\documentclass{LMCS}

\usepackage{enumerate}
\usepackage{hyperref}

\usepackage{latexsym,amsmath,amssymb}

\usepackage{prooftree}
\usepackage{diagrams}

\newcommand{\GAP}{\hspace*{.5cm}}

\newcommand{\eqdef}{=_{\mathrm{def}}}

\newcommand{\PE}{\text{PE}}

\newcommand{\comptype}[1]{\underline{#1}}
\newcommand{\TVX}{X}
\newcommand{\TVY}{Y}
\newcommand{\TVZ}{Z}

\newcommand{\algX}{\comptype{X}}
\newcommand{\algY}{\comptype{Y}}
\newcommand{\algZ}{\comptype{Z}}

\newcommand{\TA}{\comptype{\mathsf{A}}}
\newcommand{\TB}{\comptype{\mathsf{B}}}
\newcommand{\TC}{\comptype{\mathsf{C}}}
\newcommand{\TS}{\mathsf{A}}
\newcommand{\TT}{\mathsf{B}}
\newcommand{\TU}{\mathsf{C}}

\newcommand{\lfun}{\multimap}
\newcommand{\All}[2]{\forall #1.\: #2}
\newcommand{\Exists}[2]{\exists #1.\: #2}
\newcommand{\Mu}[2]{\mu #1.\: #2}
\newcommand{\Nu}[2]{\nu #1.\: #2}
\newcommand{\bang}[1]{{!{\, #1}}}
\newcommand{\ftv}[1]{\mathrm{ftv}(#1)}

\newcommand{\one}{{1}}
\newcommand{\algplus}{\oplus}
\newcommand{\algtimes}{\times^{\!\circ}}
\newcommand{\algone}{1^{\circ}}
\newcommand{\algzero}{0^{\circ}}

\newcommand{\algMu}[2]{\mu^\circ #1.\: #2}
\newcommand{\algNu}[2]{\nu^\circ #1.\: #2}

\newcommand{\algexists}[2]{\exists^{\circ} #1.\, #2}

\newcommand{\In}[2]{#1 \colon  \! #2}
\newcommand{\rIn}[2]{#1 \colon  #2}
\newcommand{\aj}[4]{#1 \mid  \! #2 \, \vdash \, \rIn{#3}{#4}}
\newcommand{\ajth}[5]{#1 \mid  \! #2 \, \vdash_{#5} \, \rIn{#3}{#4}}
\newcommand{\tj}[3]{\aj{#1}{{-}}{#2}{#3}}

\newcommand{\llambda}{\lambda^{\!\circ}}

\newcommand{\lam}[3]{\lambda \In{#1}{#2}.\: #3}
\newcommand{\lamnt}[2]{\lambda #1.\: #2}
\newcommand{\llam}[3]{\llambda \In{#1}{#2}.\: #3}
\newcommand{\llamnt}[2]{\llambda #1.\: #2}
\newcommand{\Let}[3]{\mathrm{let}\: {\bang {#1}}\:\mathrm{be}\:{#2} \;\mathrm{in}\: #3}
\newcommand{\Lam}[2]{\Lambda #1.\: #2}

\newcommand{\Iff}{\Leftrightarrow}
\newcommand{\inv}[1]{{#1}^{-1}}

\newcommand{\SA}{A}
\newcommand{\SB}{B}
\newcommand{\SC}{C}

\newcommand{\algA}{\underline{A}}
\newcommand{\algB}{\underline{B}}
\newcommand{\algC}{\underline{C}}

\newcommand{\Sets}{\mathbf{Set}}
\newcommand{\CatA}{\mathcal{A}}
\newcommand{\CatC}{\mathcal{C}}

\newcommand{\SetA}{\mathbf{A}}
\newcommand{\SetC}{\mathbf{C}}

\newcommand{\algiso}{\cong^\circ}

\newcommand{\SubC}[1]{\mathrm{Sub}_{\CatC}(#1)}

\newcommand{\SubA}[1]{\mathrm{Sub}_{\CatA}(#1)}

\newcommand{\incl}{\hookrightarrow}
\newcommand{\mono}{\rightarrowtail}

\newcommand{\R}{\mathcal{R}}

\newcommand{\rhoR}{{\rho_{\!\R}}}

\newcommand{\RC}{\mathcal{R}_{\CatC}}

\newcommand{\RA}{\mathcal{R}_{\CatA}}

\newcommand{\RelC}[1]{\RC(#1)}

\newcommand{\RelA}[1]{\RA(#1)}

\newcommand{\Sem}[1]{[\! [ #1 ] \! ]}
\newcommand{\Sems}[2]{\Sem{#2}_{#1}}

\newcommand{\SemC}[2]{\CatC\Sems{#1}{#2}}
\newcommand{\SemCR}[2]{\R\Sems{#1}{#2}}

\newcommand{\SemA}[2]{\CatA\Sems{#1}{#2}}

\newcommand{\Grpd}[2]{\mathrm{gpd}\Sem{#1}(#2)}

\newcommand{\Gpd}[4]{\Grpd{#2}{\id_{#1}[#3/#4]}}

\newcommand{\Graph}[1]{\langle #1 \rangle}

\newcommand{\oprel}[1]{{#1}^{\mathrm{op}}}

\newcommand{\bor}{\mathrm{or}}
\newcommand{\CatAnd}{\CatA_{\text{nd}}}

\newcommand{\ev}{\mathrm{ev}}
\newcommand{\proj}[1]{p_{#1}} 

\newcommand{\pair}[2]{\langle #1, #2 \rangle}

\newcommand{\myvec}[1]{\mathbf{#1}}

\newcommand{\admcl}[1]{{#1}^\circ}

\newcommand{\im}{\mathrm{im}}
\newarrow{Multi}----o
\newarrow{Eq}=====

\newcommand{\braise}{\mathrm{raise}}
\newcommand{\bhandle}{\mathrm{handle}}
\newcommand{\CatAexc}{\CatA_{\text{exc}}}
\newcommand{\id}{\mathrm{id}}

\def\doi{5 (3:7) 2009}
\lmcsheading%
{\doi}
{1--31}
{}
{}
{Jul.~16, 2008}
{Aug.~\phantom{0}9, 2009}
{} 

\begin{document}

\title[Relational Parametricity for Computational Effects]{Relational Parametricity for Computational Effects}

\author[R.~E.~M{\o}gelberg]{Rasmus Ejlers M{\o}gelberg\rsuper a}
\address{{\lsuper a}IT University of Copenhagen, Denmark}
\email{mogel@itu.dk}
\thanks{{\lsuper a}Research supported by EPSRC and the Danish Agency for Science, Technology and Innovation.} 

\author[A.~Simpson]{Alex Simpson\rsuper b}
\address{{\lsuper b}LFCS, School of Informatics \\ University of Edinburgh, Scotland, UK}
\email{Alex.Simpson@ed.ac.uk}
\thanks{{\lsuper b}Research supported by an EPSRC Advanced Research Fellowship.}


\begin{abstract}
\noindent
According to Strachey, a polymorphic program is parametric if it applies
a uniform algorithm independently of the 
type instantiations
at which it is applied. The notion of relational parametricity, introduced
by Reynolds, is one possible mathematical formulation of this idea.
Relational parametricity provides a powerful tool for establishing data
abstraction properties, proving equivalences of datatypes, and establishing
equalities of programs. Such properties have been well studied in a pure
functional setting. Many programs, however, exhibit computational
effects, and are not accounted for by the standard theory of
relational parametricity. In this paper, 
we develop a foundational framework for extending the notion of
relational parametricity to programming languages with effects.
\end{abstract}

\maketitle

\section{Introduction}

\label{section:introduction}

The theory of \emph{relational parametricity}, proposed by Reynolds~\cite{reynolds:parametricity},
provides a powerful framework for establishing properties of polymorphic programs
and their types. 
Such properties include the ``theorems for free'' of Wadler~\cite{wadler:free}, 
universal properties for datatype encodings,
and representation independence properties for abstract datatypes.
These results are well established, see e.g.~\cite{plotkin-abadi:logic},
for the pure Girard/Reynolds
second-order $\lambda$-calculus (a.k.a.\ system F)
which provides 
a concise yet  remarkably
powerful 
calculus of typed total functions.

The generalisation of relational parametricity to richer calculi 
can be problematic. 
Even the addition of recursion (hence nontermination) 
causes difficulties, since the fixed-point property of recursion
is incompatible with 
certain consequences of 
relational parametricity as usually formulated.\footnote{Relational parametricity implies types form a cartesian closed category with
finite sums, and any such category with fixed points is trivial.}
This issue led Plotkin~\cite{plotkin:type-theory-recursion} to propose using 
second-order linear type theory as a framework for
combining parametricity and recursion, an idea which has since been
developed in an operational setting in~\cite{bpr:lily} and
in a denotational setting in~\cite{Mogelberg:lapl:journal}.
One of the many good properties of the resulting theory
of linear parametricity
is that it supports a rich collection of
polymorphic datatype encodings with the desired universal properties
following from relational parametricity.

The addition of recursion is just one possible 
extension of second-order $\lambda$-calculus.
In~\cite{Hasegawa:LMCS06}, M.\ Hasegawa
develops a syntactic account
of relational parametricity for an orthogonal extension obtained
by adding  control operators (such an extension was first introduced
by Parigot~\cite{parigot:sn-second-order} for proof-theoretic purposes).
An intriguing fact he observes is that, even though the
technical frameworks
for the two approaches are 
quite different, there are striking analogies
between his ``focal'' parametricity and Plotkin's linear
parametricity. Accordingly, Hasegawa
poses the question of whether it is possible to find
a unifying framework for relational parametricity 
that includes both his work and Plotkin's
linear parametricity as special cases. 

In this paper we provide a general theory of  relational
parametricity for computational effects, which answers 
Hasegawa's question in the affirmative. Not only does our approach
generalise both Plotkin's and Hasegawa's, but it also applies
across the full range of computational effects
(e.g., nondeterminism, probabilistic choice, input/output, side effects, exceptions, etc.).

We build on the work of Moggi~\cite{Moggi:89,moggi:monads-journal},
who proposed incorporating effects into type theory by
adding a new type constructor for typing ``computations'' 
rather than values.
For every type $\TT$,
one has a new type $\bang{\TT}$
(our non-standard notation is justified in Section~\ref{section:monadic})
whose elements represent computations that 
(potentially) return values in $\TT$, and which (possibly)  perform
effects along the way.
Semantically, $\bang{}$ is interpreted using a 
\emph{computational monad} that encapsulates the
relevant kinds of effect.

In order to obtain an account of relational parametricity for monads,
one  needs to solve a problem. Basic to relational parametricity is
the idea of treating types as relations. Polymorphic functions
are required to preserve derived relations under all possible
instantiations of relations to type variables. To extend this to
computational effects it is necessary to determine how
the operation $\bang{}$ determines a relation $\bang{R} \subseteq
\bang{\TS} \times \bang{\TT}$ from any relation
${R} \subseteq {\TS} \times {\TT}$. That is, one needs a ``relational lifting''
of the $\bang{}$ operation. 
The literature contains two approaches to 
defining such a relational lifting for 
$\bang{}$~\cite{g-lln,katsumata:top-top} 
(although neither is presented in the context of polymorphism).
Rather than choosing between these approaches, 
we instead side-step the issue in a surprising way:
we show that, given the right choice of underlying type theory,  $\bang{}$ 
is polymorphically definable in terms of more basic primitives
whose
relational interpretations are immediately apparent.

Our type theory, which we call $\PE$, 
is presented in
Section~\ref{section:pcat}. It is closely related to
Levy's 
system of \emph{call by push-value (CBPV)}~\cite{levy:cbpv}, which  subsumes
call-by-name and call-by-value calculi with effects.
Levy, following the lead of Filinski~\cite{filinski:thesis},
emphasises the importance of having two general classes of types:
\emph{value types}, which classify ``values'', 
 and \emph{computation types}, 
which classify ``computations''.
The intuitive difference between the two is that
``a value \emph{is}'' and ``a computation \emph{does}''.
Technically, this intuition is supported by the vast range
of semantic and operational interpretations of the framework, 
see~\cite{levy:cbpv}.

With general computation types at hand, one can give 
the $\bang{}$ constructor the following polymorphic definition:
\begin{align}
\label{equation:central}
\bang{\TT} \; & \eqdef \; \forall \algX.\; (\TT \to \algX) \to \algX 
& & \text{($\algX$ not free in $\TT$),}
\end{align}
where importantly the type variable $\algX$ ranges over 
computation types only. 
As we shall see, the type constructors used
in the definition all have natural relational interpretations, and hence
the defined $\bang{}$ operation
inherits an induced relational lifting.

In order to reason about parametricity in $\PE$, we 
build a  relationally parametric
model of our calculus. Even in the case of ordinary second-order
$\lambda$-calculus, the construction of parametric models is a nontrivial
task.
In our case, the interaction between value and
computation types contributes significant additional complexity.
To keep things as simple as possible, we work with a set-theoretic model,
exploiting the fact that it is consistent to do so
if one keeps to intuitionistic reasoning.
The details are presented in 
Sections~\ref{section:semantic-setting} 
and~\ref{section:type-interpretation}.
As a first application of the model, we prove in
Section~\ref{section:monadic}
that the 
$\bang{}$ operator, as defined by~(\ref{equation:central}) above,
does indeed enjoy its expected universal property (Theorem~\ref{thm:free:alg}).

In Section~\ref{section:special}, we consider how to specialise
the generic calculus $\PE$ to specific effects of interest.
One useful form of specialisation recurs in many examples.
It is common for effects to have associated 
operations that trigger and/or react to
``effectful'' behaviour. 
Typically, one would like to give an $n$-ary such operation 
the polymorphic type:
\begin{equation}
\label{equation:approximate}
\forall X. \;\; (\bang{\!X})^n \, \to \, \bang{\!X} \enspace .
\end{equation}
For example,
a binary nondeterministic choice operation
forms a computation by choosing between two possible
continuation computations.
Also, the ``handle''
operation for an exception $e$, can be viewed as a binary
operation where $\bhandle^e(p,q)$ behaves like $p$
unless $p$ raises exception $e$, in which case 
$q$ is executed. Since such operations are 
computed in a type-independent way, they 
are  ``{parametric}'' in the 
informal sense of Strachey. 
We show that
such operations are also parametric according to
our theory of relational parametricity.
This involves two technical developments, each of interest in its own
right.
The first relates to recent work by Plotkin and Power~\cite{plotkin-power:overview},
in which they observe that many operations on
effects are ``algebraic operations'' in the sense
of universal algebra. 
As Theorem~\ref{theorem:generic}, 
we obtain that  $n$-ary algebraic operations are in
one-to-one correspondence with (parametric) elements of type:
\begin{equation}
\label{equation:exact}
\forall \algX. \; {\algX}^n\, \to \, {\algX} \enspace ,
\end{equation}
where again $\algX$ ranges over computation types. Thus
algebraic operations can be incorporated within $\PE$ as 
constants of the above type (which is 
more informative than~(\ref{equation:approximate}), since
monadic types $\bang{\TT}$ are always computation types).

Not all useful operations on effects arise as algebraic
operations; e.g., exception handling is a counterexample.
However, exception handling can be added to $\PE$ using a
different strengthening of  (\ref{equation:approximate}) for its type:
\begin{equation}
\label{equation:linear}
\forall X. \;\; (\bang{\!X})^2 \, \lfun \, \bang{\!X} \enspace .
\end{equation}
This is indeed a strengthening of (\ref{equation:approximate}) because
the lollipop  can be understood as
restricting  the full function space to a subclass of ``linear'' 
(in a sense to be explained in the sequel) functions.
This correctness of the above typing is again based on
a general result (Theorem~\ref{theorem:lin-gen})
which characterises the (parametric) elements of the above type
in terms of a naturality condition.

In Section~\ref{section:derived}, we 
outline 
the relationship between $\PE$ and other approaches to parametricity and effects.
Plotkin's linear parametricity arises as
a specialisation of $\PE$ valid 
in the special case of ``commutative'' monads.
We also briefly discuss how 
Hasegawa's account of parametricity and control
arises as a specialisation of $\PE$.
The details for this appear
in a companion paper~\cite{mogelberg:simpson:mfps07}.
Finally, in Section~\ref{section:applicability}, we discuss
how the theory established in this paper might be applied
to derive operational properties of polymorphic languages
with effects.

\section{A polymorphic calculus}
\label{section:pcat}

\begin{figure*}[t]
\begin{gather*}
\prooftree
\justifies
\tj{\Gamma,\, \In{x}{\TT}}{x}{\TT}
\endprooftree
\GAP 
\GAP
\prooftree
\aj{\Gamma,\,\In{x}{\TT}}{\Delta}{t}{\TU}
\justifies
\aj{\Gamma}{\Delta}{\lam{x}{\TT}{t}}{\TT \to \TU}
\endprooftree
\GAP 
\GAP
\prooftree
\aj{\Gamma}{\Delta}{s}{\TT \to \TU} 
  \GAP
\tj{\Gamma}{t}{\TT} 
\justifies
\aj{\Gamma}{\Delta}{s(t)}{\TU}
\endprooftree
\\
\\
\prooftree
\aj{\Gamma}{\Delta}{t}{\TT}
\justifies
\aj{\Gamma}{\Delta}{\Lam{\TVX}{t}}{\All{\TVX}{\TT}}
\using
\, \TVX \not\in \ftv{\Gamma,\Delta}
\endprooftree
\GAP 
\GAP
\prooftree
\aj{\Gamma}{\Delta}{t}{\All{\TVX}{\TT}} 
\justifies
\aj{\Gamma}{\Delta}{t(\TS)}{\TT[\TS/\TVX]}
\endprooftree
\\
\\
\prooftree
\justifies
\aj{\Gamma}{\In{x}{\TA}}{x}{\TA}
\endprooftree
\GAP
\GAP
\prooftree
\aj{\Gamma}{\In{x}{\TA}}{t}{\TB}
\justifies
\tj{\Gamma}{\llam{x}{\TA}{t}}{\TA \lfun \TB}
\endprooftree
\GAP 
\GAP
\prooftree
\tj{\Gamma}{s}{\TA \lfun \TB} 
  \GAP
\aj{\Gamma}{\Delta}{t}{\TA} 
\justifies
\aj{\Gamma}{\Delta}{s(t)}{\TB}
\endprooftree
\\
\\
\prooftree
\aj{\Gamma}{\Delta}{t}{\TT}
\justifies
\aj{\Gamma}{\Delta}{\Lam{\algX}{t}}{\All{\algX}{\TT}}
\using
\, \algX \not\in \ftv{\Gamma,\Delta}
\endprooftree
\GAP 
\GAP
\prooftree
\aj{\Gamma}{\Delta}{t}{\All{\algX}{\TT}} 
\justifies
\aj{\Gamma}{\Delta}{t(\TA)}{\TT[\TA/\algX]}
\endprooftree
\end{gather*}
\vspace*{-.4cm}
\caption{Typing rules.}
\label{figure:typing}
\end{figure*}

We start by defining the type theory $\PE$ for polymorphism and effects. 
As discussed in the introduction, following~\cite{levy:cbpv}, $\PE$ contains
both \emph{value types} $\TS,\TT,\TU, \dots$ and \emph{computation types} $\TA,\TB,\TC, \dots$.
A central feature of our type theory is that we  allow
polymorphic type quantification over both value types and computation types.
Accordingly, 
we use $\TVX, \TVY, \TVZ, \dots$ to range over a countable set
of value-type variables, and $\algX, \algY, \algZ, \dots$ to range over
a disjoint countable set of computation-type variables.
Value types and computation types are then mutually defined by:
\begin{align*}
\TS \; & ::= \; \TVX \mid \TT \to \TU \mid \All{\TVX}{\TT} \mid
              \algX \mid \All{\algX}{\TT} \mid \TA \lfun \TB \\
\TA \; & ::= \; \TT \to \TA \mid \All{\TVX}{\TA} \mid
              \algX \mid \All{\algX}{\TA} 
\end{align*}
Note that the computation types form a 
subcollection of the value  types. The intuition here is that
any (active) computation has a corresponding (static) value, its ``thunk''.
In contrast to~\cite{levy:cbpv}, we make this passage from
computations to values syntactically invisible. 

For semantic intuition, one can think
of value types as representing sets, and of computation types
as representing Eilenberg-Moore algebras for some computational  monad on sets.
Then \mbox{$\TT \to \TU$} is the set of all functions.
The special case $\TT \to \TA$ is a computation type
because algebras are closed under powers, with the algebra
structure defined pointwise. The type $\TA \lfun \TB$ represents
the set of all algebra homomorphisms from $\TA$ to $\TB$. In general,
there is no natural algebra structure on this set, hence the
type $\TA \lfun \TB$  is not a computation type.
Finally $\All{\TVX}{\TT}$ and $\All{\algX}{\TT}$ are
polymorphic types, with the polymorphism ranging
over value types and computation types respectively.
In either case, when $\TT$ is a computation type,
the polymorphic type is again a computation type. This
is justified by Proposition~\ref{proposition:types-well-defined} below.

Our types, which are based on function spaces and
polymorphism, are not directly comparable with Levy's~\cite{levy:cbpv},
which include sums and products. Nonetheless, we shall see in
Section~\ref{section:derived} that we can encode 
Levy's calculus within ours. Given this, our calculus 
extends Levy's  with polymorphic types (cf.\ \cite[\S12.4]{levy:cbpv}) and
linear function types. 
The latter  have 
a particularly nice explanation in terms of Levy's stack-based
operational framework, within which
a value of type  $\TA \lfun \TB$ can be understood as 
a stack
turning a computation of type $\TA$ into
a computation of type $\TB$, cf.~\cite{levy:stacks}.
In our system,
linear function types will be used crucially in the 
computation-type encodings of Section~\ref{section:derived}.

Having computation types as special value types allows
us to base our type system on a single 
judgement form:
\[
\aj{\Gamma}{\Delta}{t}{\TT} \enspace ,
\]
where $\Gamma$ and $\Delta$ are disjoint contexts of variable typings
subject to the
following conditions: either (i) $\Delta$ is empty, or (ii)
$\TT$ is a computation type and
$\Delta$ has the form 
$\In{x}{\TA}$, where $\TA$ is also a computation type.
Thus the context $\Delta$, which, following~\cite{girard:nc,girard}, we call the \emph{stoup} of
the typing judgement, 
contains at most
one typing assertion.
When we want to be explicit about which of (i) or
(ii) applies, we write:
\[
\begin{array}{ll}
\text{(i)} & \tj{\Gamma}{t}{\TT} \\
\text{(ii)} & \aj{\Gamma}{\In{x}{\TA}}{t}{\TB} \enspace .
\end{array}
\]
In the first case, the intuitive interpretation of $t$ is as an
arbitrary function from the product of all types in $\Gamma$ to
the type $\TT$. In the second case, the interpretation
of $t$ is as a function from $\Gamma \times \TA$ to $\TB$
that is an algebra homomorphism in its right-hand argument (i.e.,\
for every fixed set of values for the $\Gamma$ variables, the induced
function from $\TA$ to $\TB$ is a homomorphism). From this
interpretation, one sees why the stoup is restricted to computation
types, and also why, when the stoup is nonempty, the 
result type is required to be a computation type.

The type system is presented in Figure~\ref{figure:typing}.
The side conditions refer to the set $\ftv{\Gamma}$ of free type variables
in a context $\Gamma$, which is defined in the obvious way.
Of course, the type rules are restricted to apply only when
the premises satisfy the conditions on judgements imposed
above. 
In such cases, the 
rule conclusions also satisfy these conditions.

The following simple lemmata state basic properties of the type system.

\begin{lem}[Unicity of types]
For any $\Gamma, \Delta, t$ there is at most one type $\TT$
such that $\aj{\Gamma}{\Delta}{t}{\TT}$.
\end{lem}

\begin{lem}[Substitution]
\leavevmode
\begin{enumerate}[\em(1)]
\item If $\aj{\Gamma,\, \In{x}{\TS}}{\Delta}{t}{\TT}$ and $\tj{\Gamma}{s}{\TS}$
then $\aj{\Gamma}{\Delta}{t[s/x]}{\TT}$.

\item If $\aj{\Gamma}{\In{x}{\TA}}{t}{\TB}$ and $\aj{\Gamma}{\Delta}{s}{\TA}$
then $\aj{\Gamma}{\Delta}{t[s/x]}{\TB}$.
\end{enumerate}
\end{lem}

\proof
Both statements are proved by induction over the depth of the typing derivation for $t$. For example, consider the second statement in the case of $t = u \, u'$, where \mbox{$\aj{\Gamma}{\In{x}{\TA}}{u}{\TU \to\TB}$} and $\tj{\Gamma}{u'}{\TU}$. In this case $(u\, u')[s/x] = u[s/x]\,u'$ and by induction hypothesis \mbox{$\aj{\Gamma}{\Delta}{u[s/x]}{\TU \to \TB}$}, so $\aj{\Gamma}{\Delta}{u[s/x]\, u'}{\TB}$. 
\qed

\begin{figure}
\begin{align*}
\one  & \eqdef  \All{\TVX}{\TVX \to \TVX } \\
\TS \times \TT   & \eqdef  \All{\TVX}{(\TS \to \TT \to \TVX) \to \TVX} 
 & & \!\!\text{($\TVX \! \not\in \! \ftv{\TS,\!\TT}$)} \\
0  & \eqdef  \All{\TVX}{\TVX} \\
\TS +  \TT   & \eqdef  \All{\TVX}{(\TS\! \to\! \TVX) \to (\TT \!\to \!\TVX) \to \TVX} \!\!\!
 & & \!\!\text{($\TVX \! \not\in \!\ftv{\TS,\!\TT}$)} 
& 
 \\
\Exists{\TVX}{\TT} \: & \eqdef \: \All{\TVY}{(\All{\TVX}{(\TT \to \TVY})) \to \TVY} 
 & & \text{($\TVY \not\in \ftv{\TT}$)}\\
\Mu{\TVX}{\TT} \: & \eqdef \: \All{\TVX}{(\TT \to \TVX) \to \TVX} & & \text{($\TVX$ +ve in $\TT$)} \\
\Nu{\TVX}{\TT} \: & \eqdef \: \Exists{\TVX}{(\TVX \to \TT) \times  \TVX} & & \text{($\TVX$ +ve in $\TT$)} 
 \\
 \Exists{\algX}{\TT} \: & \eqdef \: \All{\TVY}{(\All{\algX}{(\TT \to \TVY})) \to \TVY} 
 & & \text{($\TVY \not\in \ftv{\TT}$)}
\end{align*}
\vspace*{-.7cm}
\caption{Definable value types}
\label{figure:definable}
\end{figure}

It is immediate that the type system for value types extends the
standard second-order $\lambda$-calculus of Girard and Reynolds.
Indeed, the typing rules for the relevant types ($\TVX$, $\;\TT \to \TU$ and
$\All{\TVX}{\TT}$), when restricted to the case with empty stoup, are just
the usual ones. 
It is well-known that the second-order 
$\lambda$-calculus is powerful enough to encode many
type constructors including products, sums, inductive
and coinductive types. We include those definitions
we shall need later in Figure~\ref{figure:definable}. 
These encodings are all standard apart from the last one
which is existential quantification over computation types. The 
introduction and elimination constructs for the definable value types 
are encoded in most cases as in the second-order $\lambda$-calculus,
but the presence of the stoup in $\PE$ means that in some cases a slight variation of 
these encodings must be used.
A more detailed discussion of this issue appears 
in~\cite[Sec.\ 4]{mogelberg:simpson:logic}.

\section{Semantic setting}
\label{section:semantic-setting}

In the previous section, we appealed to semantic intuition by
explaining value types as sets and computation types as algebras
for a monad on sets.
Unfortunately, this intuition runs into the technical problem
that there are no set-theoretic models of polymorphism~\cite{reynolds:set-theoretic}.
However, it was shown by Pitts~\cite{pitts:set-theoretic} that set-theoretic models
of polymorphism are possible if \emph{intuitionistic} set theory
is used rather than ordinary \emph{classical} set theory. 
We shall exploit this by working with such an
intuitionistic set-theoretic model.
The advantage of this strategy is
that the set-theoretic framework allows the development to concentrate entirely
on the difficulties inherent in defining
a suitable notion of relational parametricity, which
are formidable in themselves, rather than on
incidental details specific to a particular concrete  model.
Our approach results in no loss of generality.
All denotational models of relational parametricity of which
we are aware can be exhibited as full subcategories of 
models of intuitionistic set theory.

The intuitionistic set theory we use in this paper is Friedman's
Intuitionistic Zermelo-Fraenkel set theory (IZF), which is
the established intuitionistic counterpart of
classical Zermelo-Fraenkel set theory (ZF). The theory
IZF is axiomatized  over intuitionistic first-order
logic with equality. 
The axioms of IZF are the usual axioms of classical ZF, except that
Collection is taken as an axiom schema instead of Replacement, and 
Foundation is formulated as a principle of transfinite induction
over the membership relation. One reason for assuming the Collection schema
is that it is strictly stronger than  Replacement 
under intuitionistic logic. The
reformulation of Foundation is required because the usual
versions of the axiom imply the Law of Excluded Middle (LEM), whence
classical logic. (The Axiom of Choice
also implies LEM, and so is not considered.) The naturalness of IZF is
underlined by the existence of a wide range of
Kripke, sheaf and realizability models. 
For a detailed summary of the axioms and properties of
IZF, see \v{S}\v{c}edrov's survey article~\cite{scedrov:ist}.

Henceforth in this paper, we use IZF as our 
mathematical meta-theory.
To keep matters readable, we 
work \emph{informally} within IZF, 
just as in ordinary mathematical practice
one works informally in ZF.  This approach
is deliberately chosen to avoid cluttering the mathematics
of the arguments with the formalities of the metatheory.
(Nevertheless, when it is particularly helpful to do so,
we shall occasionally remark on technical aspects of the formalization.)
In fact, to the casual reader, it will not seem that much out of the ordinary
is going on. Given the similarity between the axioms of ZF and IZF,
reasoning within IZF feels
very much like reasoning within classical ZF. 
Essentially, the only 
practical difference is that one has to adhere to
the discipline of intuitionistic logic. The reader should try to
be sensitive to this issue, because our adherence to intuitionistic
logic is essential to the consistency of this paper. Nonetheless,
since IZF is a subtheory of ZF,
readers who are not familiar with the distinctions between 
intuitionistic and classical reasoning, should anyway be able to follow the 
mathematical development.
Such readers will, however, have to place their trust in the authors that the 
reasoning principles of IZF are never violated. For anyone who 
wishes to learn more about reasoning in intuitionistic set theory,
a good starting place is~\cite{aczel-rathjen}.

As is common in set-theoretic reasoning, we shall sometimes have to work with
collections of sets that are too ``large'' to themselves form a set; that is,
with proper classes. When working with IZF (as with classical ZF), classes
are accommodated by taking them as being
represented by formulas: a formula $\phi$ 
with distinguished free variable $x$ represents the class
$\{x \mid \phi\}$. In practice, it would be a nuisance to
always have to work with concrete formulas $\phi$. Instead, 
we shall typically say: ``let $X$ be a class
then\ \dots'', without specifying a particular formula $\phi$ that
represents $X$. Such reasoning can be understood schematically
as being valid relative to any possible formula instantiating $X$
(and, in practice, there may be several different concrete
instantiations that satisfy all assumed properties of $X$). Alternatively,
it is possible to view the development as taking place in an extension 
of the language of set theory with a new unary predicate for every assumed class. This latter
viewpoint is slightly more general, since, in models,  it allows classes to
be collections other than  those specified by formulas
in the language of set theory. Such mild added generality 
is natural  if one interprets our reasoning in
the categorical models of IZF given by 
\emph{algebraic set theory}~\cite{joyal-moerdijk,simpson:lics99},
where the category of classes is the primary category of interest, and
class predicates can be interpreted as objects in such a category.
Whichever viewpoint one takes on whether one thinks of the language as extended with
class predicates or not, the underlying set theory remains ``morally'' unchanged,
and we shall accordingly continue to refer to it as IZF.

We now begin the technical development within IZF.
As discussed above, value types will be modelled as sets. However,
it is known that
it is not possible to interpret types in the second-order $\lambda$-calculus
as arbitrary sets~\cite{pitts:power-types}.
Thus we require a collection of special
sets for interpreting types. Such special sets
need to be closed under the 
set-theoretic operations used in the interpretation.
Accordingly, we assume that we have a full subcategory $\CatC$ of the
category $\Sets$ of sets that satisfies:

\begin{enumerate}[\hbox to8 pt{\hfill}]
\item{\hskip-12 pt\bf (C1):}\ \label{C:b} For any set-indexed family $\{\SA_i \}_{i \in I}$ of sets in $\CatC$,
the set-theoretic product $\prod_{i \in I} \SA_i$ is again in $\CatC$.

\item{\hskip-12 pt\bf (C2):}\ \label{C:c} Given $\SA, \SB \in \CatC$ and functions $f,g \colon \SA \to \SB$,
the equalizer $\{x \in \SA \mid f(x) = g(x)\}$ is again in $\CatC$.
\end{enumerate}
In other words, the category
$\CatC$ is small-complete with limits inherited from $\Sets$. Since
function spaces are powers, for any set $\SA$ and any $\SB \in \CatC$, the
function space $\SB^\SA$ is in $\CatC$, i.e.,
$\CatC$ is an \emph{exponential ideal} of $\Sets$.
In particular, $\CatC$ is cartesian closed.
In addition, we require:
\begin{enumerate}[\hbox to8 pt{\hfill}]
\item{\hskip-12 pt\bf (C3):}\ \label{C:d} There is a set $\SetC$ of objects of $\CatC$ such that,
for any $\SA \in \CatC$, there exists $\SB \in \SetC$ with $\SB \cong \SA$.

\item{\hskip-12 pt\bf (C4):}\ \label{C:a} If $\SA \in \CatC$ and $\SA \cong \SB$ in $\Sets$ then $\SB \in \CatC$.
\end{enumerate}
These two properties pull in opposite directions. Property~(C3) 
requires that $\CatC$ enjoys a smallness constraint, which will be  used
to interpret polymorphism. Explicitly, (C3) says that 
$\CatC$ is weakly equivalent to its small full subcategory
on the set of objects $\SetC$. It is not, however, a small category itself, since~(C4) forces 
$\CatC$ to have a proper class of objects.

In classical set theory, conditions (C1) and (C3) together imply that
every object in $\CatC$ is either the empty set or a singleton
set (cf.\ Freyd's argument that a weakly small category with small products
is a preorder, see~\cite[Proposition V.2.3]{macLane}).
The reason we work in IZF is that
this renders it consistent for there to be a nontrivial category
satisfying all of (C1)--(C4). 
Indeed, it is consistent for the natural numbers to be an
object of $\CatC$.
This consistency property derives from 
the work of Hyland \emph{et.\ al.}~on small-complete small categories~\cite{hyland:small-complete,hrr:discrete}.
However, our perspective is slightly different. 
Rather than assuming
a small category that is complete only in a restricted technical
sense~\cite{hrr:discrete,robinson:hcip}, our category $\CatC$ is 
assumed to be genuinely complete,
but only weakly equivalent to a small category.
This approach, which is taken from~\cite{rosolini-simpson:strictness},
offers several conveniences. For example, it allows us to assume (C4),
which, as well as being a natural repleteness condition on $\CatC$,
makes it easy to show that sets we have defined 
explicitly are actually in $\CatC$.

According to our informal explanation of computation types in
Section~\ref{section:pcat}, they can be interpreted as 
Eilenberg-Moore algebras for a  monad $T$ on $\CatC$.
For any such monad $T$, the category $\CatA$ of algebras comes with
a forgetful functor $U \colon \CatA \to \CatC$ and the
following properties are satisfied.
\begin{enumerate}[\hbox to8 pt{\hfill}]
\item{\hskip-12 pt\bf (A1):}\ $U$ ``weakly creates limits'' in the following sense. For
  every diagram $\Delta$ in $\CatA$ and limiting cone $\lim(U(\Delta))$
  of $U(\Delta)$ in $\CatC$, there exists a specified\footnote{By a 
  \emph{specified} limiting cone we mean that we are given a (class) function 
  that maps any diagram $\Delta$ and limiting cone for $U(\Delta)$
  to the required limiting cone in $\CatA$.} limiting cone
  $\lim \Delta$ of $\Delta$ in $\CatA$ such that $U(\lim \Delta) =
  \lim(U(\Delta))$.

\item{\hskip-12 pt\bf (A2):}\ $U$ reflects isomorphisms (i.e., if $U\! f$ is an isomorphism in $\CatC$ then
$f$ is an isomorphism in $\CatA$).

\item{\hskip-12 pt\bf (A3):}\ For objects $\algA,\algB$ of $\CatA$, the hom-set $\CatA(\algA,\algB)$ is an
object of $\CatC$.

\item{\hskip-12 pt\bf (A4):}\ There exists a set $\SetA$ of objects of $\CatA$ such that for every $\algA \in \CatA$, there exists $\algB \in \SetA$ with $\algB$ isomorphic to $\algA$.
\end{enumerate}
\begin{lem}
Suppose $\CatC$ satisfies (C1)--(C4) and let $T$ be a monad on $\CatC$. Then the category $\CatA$ of Eilenberg-Moore 
algebras for $T$ and the forgetful functor $U \colon \CatA \to \CatC$ satisfy (A1)--(A4).
\end{lem}

\proof
Properties (A1) and (A2) are standard, indeed the forgetful functor creates limits, which implies (A1).
Property (A3) holds because $\CatA(\algA, \algB)$ arises as an equalizer in
$\CatC$ of two evident functions $(U\algB)^{U\algA} \to (U\algB)^{TU\!\algA}$. For property (A4) define
\[\SetA  = \{(\algA, \xi) \mid \algA \in \SetC, \text{and}\
\xi \text{ is an Eilenberg-Moore algebra structure on }\algA\}.
\]
\qed

The reason for identifying (A1)--(A4) is that,
in order to interpret the calculus of Section~\ref{section:pcat},
it is sufficient to work with any category $\CatA$ and
functor $U \colon \CatA \to \CatC$ satisfying (A1)--(A4) above.\footnote{In particular, 
the weakening of limit creation in (A1) is crucial to the
application in~\cite{mogelberg:simpson:mfps07}.}
Henceforth, we assume this situation.

It is convenient to maintain
algebraic terminology for the category $\CatA$.
Thus we call the objects of $\CatA$ \emph{algebras}.
By (A1) and (A2), the functor $U$ is faithful,
thus we can identify the morphisms
$\CatA(\algA,\algB)$ with 
special functions 
from $U\!\algA$ to $U\!\algB$, which we call \emph{homomorphisms}.
We write $\algA \!\lfun\! \algB$ for the set of homomorphisms
from $\algA$ to $\algB$. (N.B.\ by (A3) the set $\algA \lfun \algB$ is an object of  $\CatC$.) The notation $\algA \algiso \algB$ means $\algA, \algB$ are isomorphic in $\CatA$.

In Section~\ref{section:type-interpretation} 
we interpret the type theory of 
Section~\ref{section:pcat} using $U \colon \CatA \to \CatC$.
In doing so, we formulate relational parametricity
using binary relations
in the categories 
$\CatC$ and $\CatA$.
As usual, these are defined as subobjects of products.
First, let us review some basic properties of subobjects in
$\CatC$ and $\CatA$.

For every object $\SA$ of $\CatC$, we write
$\SubC{\SA}$ for the
set of subobjects of $\SA$ in the category $\CatC$. Since 
the inclusion $\CatC \incl \Sets$ preserves limits and hence
monomorphisms, this 
is explicitly defined by:
\[
\SubC{\SA} \: = \: \{\SB \in \CatC \mid \SB \subseteq \SA \} \, .
\]
We call the elements of $\SubC{\SA}$ the \emph{$\CatC$-subsets} of $\SA$.

Similarly, we write 
$\SubA{\algA}$ for the
collection of subobjects of an algebra $\algA$ in $\CatA$. Because
$U$ preserves limits, every mono $\algB \mono \algA$ in $\CatA$ is mapped
by $U$ to a mono $U\algB \mono U\algA$ in $\CatC$.
Thus, for every $\algA \in \CatA$, the functor $U$ determines
a function $\SubA{\algA} \to \SubC{U\algA}$.
The lemma below shows that we can view 
subobjects of $\algA$ in $\CatA$ as special subobjects of 
$U\algA$ in $\CatC$.

\begin{lem}
\label{lemma:pr}
The function $\SubA{\algA} \to \SubC{U\algA}$ preserves and reflects the ordering.
\end{lem}

\proof
We show that it reflects the ordering. Suppose $\algB \mono \algA$ and $\algC \mono \algA$ represent subobjects of $\algA$ such that the subobject represented by $U\algB \mono U\algA$ is smaller than that represented by \mbox{$U\algC \mono U\algA$.} Then there exists an $f$ such that the square below is a pullback. 
\begin{equation}
\label{diag:pbk}
\begin{diagram}
U\algB \SEpbk & \rTo^f & U\algC \\
\dEq  & & \dTo \\
U\algB & \rTo & U \algA
\end{diagram}
\end{equation}
By (A1) there exists a pullback diagram
\begin{diagram}
\algB' \SEpbk & \rMulti & \algC \\
\dMulti  & & \dMulti \\
\algB & \rMulti & \algA ,
\end{diagram}
in $\CatA$ mapped by $U$ to (\ref{diag:pbk}), and by (A2) the map $\algB' \lfun \algB$ is an isomorphism, so $\algB \mono \algA$ represents a smaller subobject than $\algC \mono \algA$.
\qed

We say that $\SA\subseteq U\algA$ \emph{carries a subalgebra} if it represents a subobject in the 
image of the map $\SubA{\algA} \to \SubC{U\algA}$ induced by $U$.
In fact, $\SubA{\algA}$ is given explicitly by:
\[
\SubA{\algA}  =  \{ \! \SB \in \CatC  \! \mid \! \text{$\SB \subseteq U\algA$ and carries a subalgebra of $\algA$}\} \enspace .
\]

Axiom (A1) gives a way of picking representatives in $\CatA$ for subalgebras presented by subsets:
\begin{lem} \label{lem:canonical:subalgebra}
For each $\SA\in \SubA{\algA}$ there is a specified algebra $\algB$ and mono $f\colon \algB\mono \algA$ in $\CatA$ such that $Uf$ is the inclusion of $\SA$ into $U\algA$. 
\end{lem}

\proof 
Suppose $\SA\subseteq U\algA$ carries a subalgebra of $\algA$. Then the set
\begin{equation} \label{eq:subsetdiagram}
\{(\algB,i) \mid \algB \in\SetA, i\colon \algB\lfun \algA \text{ mono}, U(i) \cong (\SA\subseteq U\algA)\}
\end{equation}
where the last isomorphism is an isomorphism of subobjects, is non-empty. The set (\ref{eq:subsetdiagram}) indexes a diagram in $\CatA$, and $\SA$ is a limit in $\CatC$ of $U$ applied to this diagram. Now, (A1) gives the specified mono projecting to $\SA\subseteq U\algA$.
\qed

We introduce notation for binary relations. For $\SA \in \CatC$, we
write $\Delta_{\SA}$ for the diagonal (identity) relation in
$\SubC{\SA\times \SA}$. Similarly, for $\algA \in \CatA$, we write
$\Delta_{\algA}$ for the diagonal relation on $U\algA$, which is
indeed in $\SubA{\algA \times \algA}$. 
For $R\in \SubC{\SA \times \SB}$, we write
$\oprel{R}$ for its opposite relation in $\SubC{\SB \times \SA}$.
Similarly, for $Q\in \SubA{\algA \times \algB}$, we have
\mbox{$\oprel{Q}\in \SubA{\algB \times \algA}$.}
For $f \colon \SA' \to \SA$ and $g\colon \SB'\to \SB$ in $\CatC$,
we write $\inv{(f,g)}R$ for \mbox{$\{(x,y) \mid (f(x),
g(y)) \in R\}$.} Notice that if $f \colon \algA' \lfun \algA, g\colon \algB'\lfun
\algB$ in $\CatA$ and $Q\in \SubA{\algA \times \algB}$ then $\inv{(f,g)}Q\in
\SubA{\algA'\times \algB'}$.

To formulate relational parametricity, we require
two specified collections of \emph{admissible}
relations, one $\RelC{\SA,\SB}\subseteq\SubC{\SA \times \SB}$ on objects of
$\CatC$ and one $\RelA{\algA,\algB}\subseteq\SubA{\algA\times \algB}$ on objects of
$\CatA$. These are required to satisfy:
\begin{enumerate}[\hbox to8 pt{\hfill}]
\item{\hskip-12 pt\bf (R1):}\ For each object $\SA$ of $\CatC$ the diagonal relation $\Delta_{\SA}$ is in $\RelC{\SA,\SA}$ and likewise for each object $\algA$ of $\CatA$ the diagonal $\Delta_{\algA}$ is in $\RelA{\algA,\algA}$.
\item{\hskip-12 pt\bf (R2):}\ Admissible relations are closed under reindexing, i.e., if $R\in \RelC{\SA,\SB}$ and \mbox{$f\colon \SA'\to \SA$,} $g\colon \SB' \to \SB$, then $\inv{(f,g)}R \in \RelC{\SA', \SB'}$ and if $Q\in \RelA{\algA,\algB}$ and $f\colon \algA'\lfun \algA$, $g\colon \algB' \lfun \algB$, then $\inv{(f,g)}Q \in \RelA{\algA', \algB'}$
\item{\hskip-12 pt\bf (R3):}\  For any set of admissible $\CatC$-\ (respectively $\CatA$-)relations on the same pair of objects, 
the intersection is an admissible $\CatC$-\ (respectively $\CatA$-)relation.
\item{\hskip-12 pt\bf (R4):}\ $\RelA{\algA,\algB} \subseteq \RelC{U\algA, U\algB}$.
\end{enumerate}
\noindent
(R1)  and (R2)  imply that graphs of functions are admissible, i.e., if $f\colon \SA\to \SB$  then $\Graph f \eqdef \{(x,y) \mid f(x)=y\} \in \RelC{\SA,\SB}$  and if $g\colon \algA\lfun \algB$  then $\Graph g\in \RelA{\algA,\algB}$, for $\Graph f = \inv{(f, \id_{\SB})}\Delta_{\SB}$ and $\Graph g = \inv{(g, \id_{\algB})}\Delta_{\algB}$. Note also that if $\algA, \algB \in \CatA$ and $R\subseteq U\algA \times U\algB$ is any subset, then there exists a smallest admissible relation $\admcl{R}\in \RelA{\algA, \algB}$ containing $R$, as we may take $\admcl{R}$ to be the intersection of all admissible relations containing $R$.

In many concrete models  $\RelC{\SA,\SB} = \SubC{\SA \times \SB}$ and 
$\RelA{\algA,\algB} = \SubA{\algA\times \algB}$ will be a natural choice of admissible relations.
\begin{lem}\label{lem:Raxioms:subsets}
If $\CatC$ satisfies (C1)--(C4) and $U\colon \CatA \to \CatC$ satisfies (A1)--(A4) then the collections $\RelC{\SA, \SB} = \SubC{\SA \times \SB}$ and $\RelA{\algA, \algB} = \SubA{\algA \times \algB}$
satisfy (R1)--(R4). 
\end{lem}

\proof
We just show that $\SubA{\algA,\algB}$ is closed under intersections. So suppose we are given a set $(Q_i)_{i\in I}$ of subsets in $\SubA{\algA,\algB}$. We need to show that the subset $\bigcap_i Q_i \subseteq U\algA\times U\algB$ carries a subalgebra of $\algA\times \algB$. Denote for each $i\in I$ by $q_i \colon Q'_i \lfun \algA\times \algB$ the mono in $\CatA$ above the inclusion $Q_i \subseteq U\algA\times U\algB$ as specified by Lemma~\ref{lem:canonical:subalgebra}. Then the limit of the diagram given by the $q_i$ as weakly created by $U$ is a subalgebra of $\algA\times \algB$ above $\bigcap_i Q_i \subseteq U\algA\times U\algB$. 
\qed

By a \emph{parametric model of $\PE$} we shall mean any category $\CatC$ satisfying
(C1)--(C4), together with a category $\CatA$ and functor $U \colon \CatA \to \CatC$
satisfying (A1)--(A4) and collections $\RC$ and $\RA$ satisfying (R1)--(R4) above.
The proposition below shows that every monad on $\CatC$ gives rise to a 
parametric model of $\PE$. Thus the theory of relational parametricity
for $\PE$ that we shall develop over such models is applicable to 
arbitrary computational monads.

\begin{prop}\label{prop:alg-model}
Given $\CatC$ satisfying (C1)--(C4) and a monad $T$ on $\CatC$, let $\CatA$ be the category
of algebras for the monad, $U$ the forgetful functor and define
$\RelC{\SA, \SB} = \SubC{\SA \times \SB}$ and $\RelA{\algA, \algB} = \SubA{\algA \times \algB}$.
This data defines a parametric model of $\PE$.
\end{prop}

\proof
We have already argued above that (A1)--(A4) are satisfied, and (R1)--(R4) are satisfied by Lemma~\ref{lem:Raxioms:subsets}.
\qed

Notice that the assumption, familiar from the literature on computational
monads~\cite{Moggi:89,moggi:monads-journal},  that the 
monad $T$ is strong does not need to be included in the
above result. This is for the simple reason that our set-theoretic
setting renders all monads on $\CatC$ 
strong. For any monad $T$, one 
defines the strength $t_{\SA, \SB} \colon \SA \times T(\SB) \to T(\SA
\times \SB)$ as
\[t_{\SA, \SB} (x,y) = T(\pair{x}{-})(y)
\]
where $\pair{x}{-}\colon \SB \to \SA \times \SB$ maps $y$ to $(x,y)$.
Moreover, this strength is unique because $\CatC$ has enough points~\cite[Proposition~3.4]{moggi:monads-journal}.

Although 
Proposition~\ref{prop:alg-model}
is a useful general result, we comment
that some applications of $\PE$ require a different choice of model.
For example, the application 
of $\PE$ to control in~\cite{mogelberg:simpson:mfps07}
makes crucial use of the permitted flexibility in the 
definition of model. 
Here, we briefly describe the steps taken 
in~\emph{op.\ cit.}, in order to 
illustrate some of the variations of 
model construction available. 
The construction begins with a category
$\CatC$ satisfying (C1)--(C4), together with a chosen 
object $R$ of $\CatC$. For technical reasons (see below),
the object $R$ is used to isolate
the full subcategory  $\CatC_R$ of \emph{$R$-replete} objects
in $\CatC$, in the sense
of~\cite{hyland:first-steps}. 
Next, $\CatA$ together with $U$ are obtained 
by building $\CatA$ as a
certain carefully defined category equivalent to ${\CatC_R}^\text{op}$, and
$U$ as a functor naturally isomorphic to $R^{(-)}$. 
This situation satisfies (A1)--(A4). The interesting cases are:
(A1), which holds by the way $\CatA$ and $U$ are constructed;
and (A2), which holds because we restricted $\CatA$  to the
$R$-replete objects. Finally, whereas 
$\RelC{\SA,\SB}$ is defined to be $\SubC{\SA \times \SB}$,
it is necessary, for the application to parametricity for control,
to define $\RelA{\algA,\algB}$ to be the subset of 
$\SubA{\algA\times \algB}$ consisting of the $\top\top$-closed relations,
in the sense of Pitts~\cite{pitts:ppoe} (see also~\cite{katsumata:top-top}),
as induced by the diagonal relation $\Delta_R$ on $R$.
For full details of this construction, 
the reader is referred to~\cite{mogelberg:simpson:mfps07}.

One reason that the model construction outlined above departs from
the form of model provided by 
Proposition~\ref{prop:alg-model} is that,
although there is an underlying continuations monad
$R^{R^{(-)}}$ present, the category $\CatA$
is not in general equivalent to the category of algebras for this monad.
The usefulness of such more general situations is already familiar
from Levy's work on CBPV~\cite{levy:cbpv}, where
the natural adjunction model of control does not involve the
Eilenberg-Moore category. One of the strengths of our axiomatic framework
is that it is able to accommodate such models.

One of the drawbacks of our framework is that certain
convolutions are sometimes necessary in order to construct a model
satisfying the properties we require. For example, in the
model of control outlined
above (and described fully in~\cite{mogelberg:simpson:mfps07}),
awkward steps are taken in order to
satisfy properties (A1) and (A2). An arguably
preferable approach would be to work with the more natural
model in which $\CatA$ is simply $\CatC^\text{op}$ and
$U$ is $R^{(-)}$, as in~\cite{levy:cbpv}, even though
(A1) and (A2) are then violated. This raises the question of
whether the awkward properties (A1) and (A2) can be weakened.
We shall return to this question in 
Section~\ref{section:derived}.

\section{Interpreting the calculus}
\label{section:type-interpretation}

\begin{figure*}[t]
\begin{align*}
\SemC{\gamma}{\TVX} & = \gamma(\TVX)
\\
\SemC{\gamma}{\TT \to \TU} & =  {\SemC{\gamma}{\TU}}^{\SemC{\gamma}{\TT}}
\\
\SemC{\gamma}{\All{\TVX}{\TT}} & =
\{ \pi \in \prod_{\SA \in \SetC} \SemC{\gamma[\SA/\TVX]}{\TT} \mid \forall \SA,\SB \in \SetC,
\, \forall  R \in \RelC{\SA,\SB}. \; \SemCR{\Delta_\gamma[R/\TVX]}{\TT}(\pi_{\SA},\pi_{\SB}) \}
\\
\SemC{\gamma}{\algX} & = U (\gamma(\algX))
\\
\SemC{\gamma}{\TA \lfun \TB} & = \SemA{\gamma}{\TA} \lfun \SemA{\gamma}{\TB}
\\
\SemC{\gamma}{\All{\algX}{\TT}} & = 
\{ \kappa \in \prod_{\algA \in \SetA} \SemC{\gamma[\algA/\algX]}{\TT} \mid \forall \algA,\algB \in \SetA,
\, \forall  Q \in \RelA{\algA,\algB}. \; \SemCR{\Delta_\gamma[Q/\algX]}{\TT}(\kappa_{\algA},\kappa_{\algB}) \}
\enspace .
\\[15pt]
\SemA{\gamma}{\TT \to \TA} & = {\SemA{\gamma}{\TA}}^{\SemC{\gamma}{\TT}}
\\
\SemA{\gamma}{\All{\TVX}{\TA}} & =
\{ \pi \in {\prod_{\SA \in \SetC}}
\SemA{\gamma[\SA/\TVX]}{\TA} \mid \forall \SA,\SB \in \SetC,
\, \forall  R \in \RelC{\SA,\SB}. \; \SemCR{\Delta_\gamma[R/\TVX]}{\TA}(\pi_{\SA},\pi_{\SB}) \}
\\
\SemA{\gamma}{\algX} & = \gamma(\algX)
\\
\SemA{\gamma}{\All{\algX}{\TA}} & = 
\{ \kappa \in {\prod_{\algA \in \SetA}}
\SemA{\gamma[\algA/\algX]}{\TA} \mid \forall \algA,\algB \in \SetA,
\, \forall  Q \in \RelA{\algA,\algB}.\; \SemCR{\Delta_\gamma[Q/\algX]}{\TA}(\kappa_{\algA},\kappa_{\algB}) \}
\enspace .
\end{align*}
 \vspace*{5pt}
\begin{align*}
\SemCR{\rho}{\TVX} (x_1,x_2)  & \Iff \rhoR(\TVX)(x_1,x_2)
\\
\SemCR{\rho}{\TT \to \TU} (f_1,f_2) &  \Iff
\forall x_1 \in \SemC{\rho_1}{\TT}, x_2 \in \SemC{\rho_2}{\TT}.\;
\SemCR{\rho}{\TT}(x_1,x_2)  \implies  
\SemCR{\rho}{\TU}(f_1(x_1),f_2(x_2))
\\
\SemCR{\rho}{\All{\TVX}{\TT}} (\pi_1,\pi_2) &   \Iff
\forall \SA_1,\SA_2 \in \SetC, \forall  R \in \RelC{\SA_1,\SA_2}.\;
\SemCR{\rho[R/\TVX]}{\TT}((\pi_1)_{\SA_1},(\pi_2)_{\SA_2})
\\
\SemCR{\rho}{\algX} (x_1,x_2)  & \Iff \rhoR(\algX)(x_1,x_2)
\\
\SemCR{\rho}{\TA \lfun \TB} (h_1,h_2) &  \Iff
\forall x_1 \in \SemC{\rho_1}{\TA}, x_2 \in \SemC{\rho_2}{\TA}.\;
\SemCR{\rho}{\TA}(x_1,x_2) \implies  
\SemCR{\rho}{\TB}(h_1(x_1),h_2(x_2))
\\
\SemCR{\rho}{\All{\algX}{\TT}} (\kappa_1,\kappa_2) &   \Iff
\forall \algA_1,\algA_2 \in \SetA, \forall Q \in \RelA{\algA_1,\algA_2}. \;
\SemCR{\rho[Q/\algX]}{\TT}((\kappa_1)_{\algA_1},(\kappa_2)_{\algA_2}) \enspace .
\end{align*}
\vspace*{-.7cm}
\caption{Interpretation of Types}
\label{figure:types}
\end{figure*}

In this section we 
interpret $\PE$ in any parametric model as defined in Section~\ref{section:semantic-setting}.
As adumbrated there, a value type $\TT$ will be interpreted
as a set $\SemC{}{\TT}$ in $\CatC$, and
a computation type $\TA$ will be interpreted as an algebra $\SemA{}{\TA}$.
Since every computation type $\TA$ is also a value type, it
is given two interpretations, and we shall ensure that these
are related by  $U(\SemA{}{\TA}) = \SemC{}{\TA}$.
In order to incorporate relational parametricity, we shall also
give a second interpretation of a value type $\TT$
as an admissible $\CatC$-relation $\SemCR{}{\TT}$. 
In the special case of a computation type $\TA$, 
it will hold automatically that
$\SemCR{}{\TA}$ is also an admissible $\CatA$-relation.

Given a set of type variables $\Theta$, a
\emph{$\Theta$-environment} is a function
$\gamma$ mapping every value-type variable $\TVX \in \Theta$ to
an object $\gamma(\TVX)$ of $\CatC$, and every computation-type
variable $\algX \in \Theta$ to an object $\gamma(\algX)$ of $\CatA$.
A \emph{relational $\Theta$-environment} is a tuple
$ \rho = (\rho_1, \rho_2, \rhoR)$, where:
$\rho_1,\rho_2$ are $\Theta$-environments; for every
value-type variable $\TVX \in \Theta$,
\[\rhoR (\TVX) \in 
\RelC{\rho_1(\TVX),\rho_2(\TVX)} \enspace ;\]
and, for every computation-type variable $\algX \in \Theta$,
\[\rhoR (\algX) \in 
\RelA{\rho_1(\algX),\rho_2(\algX)} \enspace .\]

For each value type $\TT(\Theta)$ (i.e., type $\TT$ with
$\ftv{\TT} \subseteq \Theta$) and
$\Theta$-environment $\gamma$, we define an
object $\SemC{\gamma}{\TT}$ of $\CatC$;
and, for each computation type $\TA(\Theta)$ and 
$\Theta$-environment $\gamma$, we define an
object $\SemA{\gamma}{\TA}$ of $\CatA$.
Interdependently with the above, 
for each value type $\TT(\Theta)$ and
relational $\Theta$-environment $\rho$, we define an admissible
$\CatC$-relation
$\SemCR{\rho}{\TT} \in \RelC{\SemC{\rho_1}{\TT},\SemC{\rho_2}{\TT}}$.
The definitions are given in Figure~\ref{figure:types}.
In these definitions, the products and powers used in the
definition of $\SemC{\gamma}{\TT}$ are the ones in $\CatC$,
and those used in the definition of $\SemA{\gamma}{\TA}$ are
those in $\CatA$, as (weakly) created by $U$. 
We write $\Delta_\gamma$ for
the relational $\Theta$-environment that maps $\TVX$ (resp.\ $\algX$)
to $\Delta_{\gamma(\TVX)}$ (resp.\ $\Delta_{\gamma(\algX)}$).
We also use an obvious
notation for update of environments.
The algebras defined
by $\SemA{\gamma}{\All{\TVY}{\TA}}$ and $\SemA{\gamma}{\All{\algX}{\TA}}$
are the canonical algebras carried by the subsets of the product algebras.

\begin{prop}
\label{proposition:types-well-defined}
$\SemC{\gamma}{\TT}$, $\SemA{\gamma}{\TA}$ and $\SemCR{\rho}{\TT}$ are well defined
by Figure~\ref{figure:types}. Further, for every computation type $\TA$, it holds
that $\SemC{\gamma}{\TA} = U(\SemA{\gamma}{\TA})$ and 
$\SemCR{\rho}{\TA} \in \RelA{\SemA{\rho_1}{\TA},\SemA{\rho_2}{\TA}}$.
\end{prop}

\proof
The proof of well definedness is by induction over the structure of types.
We focus first on showing that the relational interpretation of types defines admissible relations. Notice first that the relation $\SemCR{\rho}{\TT \to \TU}$ can be rewritten as 
\[
\bigcap_{(x_1, x_2)\in \SemCR{\rho}{\TT}} \inv{(\ev_{x_1}, \ev_{x_2})} \SemCR{\rho}{\TU}
\]
where $\ev_{x_1}$ denotes the map from $\SemCR{\rho_1}{\TT \to \TU}$ to $\SemCR{\rho_1}{\TU}$ given by evaluation at $x_1$, and $\ev_{x_2}$ is defined likewise. For value types $\TT, \TU$ it follows that $\SemCR{\rho}{\TT \to \TU}$ is an admissible $\CatC$ relation from the induction hypothesis and (R2) and (R3). If $\TU$ is a computation type, $\TT\to \TU$ becomes a computation type and we must check that $\SemCR{\rho}{\TT \to \TU}$ is an admissible $\CatA$ relation. Since the object $\SemA{\rho_1}{\TT \to \TU}$ is defined as a product $\SemA{\rho_1}{\TU}^{\SemC{\rho_1}{\TT}}$ in $\CatA$ and the evaluation map $\ev_{x_1}$ is the projection, it is a homomorphism. So again $\SemCR{\rho}{\TT \to \TU}$ being admissible follows from the induction hypothesis and (R2), (R3).
The proof of the other induction cases are similar. 

To prove well definedness of $\SemA{\gamma}{\All{\TVX}{\TA}}$ notice first that the formula in Figure~\ref{figure:types} defines an element in $\SubA{\prod_{\SA \in \SetC} \SemA{\gamma[\SA/\TVX]}{\TA}}$ since it can be exhibited as the intersection
\begin{equation} \label{eq:intersection}
\bigcap_{\SA, \SB \in \SetC, R \in \RelC{\SA,\SB}} \inv{(\proj{\SA}, \proj{\SB})} \SemCR{\Delta_\gamma[R/\TVX]}{\TA}
\end{equation}
where $\proj{\SA}, \proj{\SB}$ are the projections from the product $\prod_{\SA \in \SetC}\SemA{\gamma[\SA/\TVX]}{\TA}$. The projections are homomorphisms since the product is taken in the category $\CatA$ and thus, since $\SemCR{\Delta_\gamma[R/\TVX]}{\TA}$ is an $\CatA$-subobject by induction hypothesis, (\ref{eq:intersection}) defines an $\CatA$-subobject.
We define $\SemA{\gamma}{\All{\TVX}{\TA}}$ to be the specified $\CatA$ object representing the subset as given by Lemma~\ref{lem:canonical:subalgebra}, thus defining $\SemA{\gamma}{\All{\TVX}{\TA}}$ up to identity and not just up to isomorphism.
\qed

We include some basic lemmata about the type interpretation 
without proof.

\begin{lem} \label{lemma:substitution}
Suppose $\gamma$ is a $\Theta$-environment and $\rho$ is a relational $\Theta$ environment. 
\begin{enumerate}[\em(1)]
\item If $\TT(\Theta, \TVX)$ and $\TS(\Theta)$ then 
\begin{align*}
\SemC{\gamma}{\TT [\TS/ \TVX]} & =  \SemC{\gamma[\SemC{\gamma}{\TS}/ \TVX]}{\TT} \\
\SemCR{\rho}{\TT [\TS/ \TVX]} & =  \SemCR{\rho[\SemCR{\rho}{\TS}/ \TVX]}{\TT} 
\end{align*}
\item If $\TT(\Theta, \algX)$ and $\TA(\Theta)$ then 
\begin{align*}
\SemC{\gamma}{\TT [\TA/ \algX]} & = \SemC{\gamma[\SemA{\gamma}{\TA}/ \algX]}{\TT} \\
\SemCR{\rho}{\TT [\TA/ \algX]} & = \SemCR{\rho[\SemCR{\rho}{\TA}/ \algX]}{\TT}
\end{align*}
\item If $\TB(\Theta, \TVX)$ and $\TS(\Theta)$ then 
\[\SemA{\gamma}{\TB [\TS/ \TVX]} = \SemA{\gamma[\SemC{\gamma}{\TS}/ \TVX]}{\TB}
\]
\item If $\TB(\Theta, \algX)$ and $\TA(\Theta)$ then 
\[\SemA{\gamma}{\TB [\TA/ \algX]} = \SemA{\gamma[\SemA{\gamma}{\TA}/ \algX]}{\TB}
\]
\end{enumerate}
\end{lem}

\begin{lem} \label{lemma:opposite}
For all types $\TS(\Theta)$ and any $\Theta$-environment $\gamma$ the relations $\oprel{\SemCR{\gamma}{\TS}}$ and $\SemCR{\oprel{\gamma}}{\TS}$ are equal, where $\oprel{\gamma} $ is the environment obtained by composing $\gamma$ with the function $\oprel{(-)}$. 
\end{lem}


\begin{lem}[Identity extension]
\label{lemma:identity}
For any type $\TT(\Theta)$ and $\Theta$-environment
$\gamma$, it holds that
$\SemCR{\Delta_\gamma}{\TT} = \Delta_{\SemC{\gamma}{\TT}}$.
\end{lem}
\noindent
The  above lemmata are all easily proved by induction on types.

The interpretations of polymorphic types have been defined by taking
products over the sets $\SetC, \SetA$ respectively, but for the
interpretation of terms below, it is crucial that we can define
projections out these products for every $\SA$ in $\CatC$
(respectively $\algB$ in $\CatA$) and not just for those objects in the
sets $\SetC, \SetA$. Essentially, we would like to be able to treat these
polymorphic types as if they had been defined using products over
the classes of objects of $\CatC$ and $\CatA$, 
even though  set theory does not allow
us to define such large products. It is a pleasing fact that
restriction to the parametric elements of the products allows us to do
just that, as the sequence of results from Proposition~\ref{prop:groupoidnew} to
Lemma~\ref{lemma:proj:hom} below establishes.
The idea essentially
goes back to~\cite{rosolini-simpson:strictness}, and was used in~\cite{Mogelberg:sdt-lapl} to construct a model of parametric polymorphism in the sense of fibered category theory. 

To formulate the first result, we define a \emph{morphism} from 
$\Theta$-environments $\gamma$ to another $\gamma'$ to be a 
family $\myvec{f}$ of functions indexed by type variables in $\Theta$
satisfying: for every value-type variable $\TVX \in \Theta$, the function
$\myvec{f}_{\TVX}$ is a function from $\gamma(\TVX)$ to $\gamma'(\TVX)$;
and, for every computation-type
variable $\algX \in \Theta$, the function
$\myvec{f}_{\algX}$ is a homomorphism from $\gamma(\algX)$ to $\gamma'(\algX)$.
Morphisms of $\Theta$-environments form a category under pointwise composition,
and a $\Theta$-environment \emph{isomorphism} is just an isomorphism in
this category. Given a $\Theta$-environment morphism $\myvec{f}$
from $\gamma$ to $\gamma'$, we write $\myvec{\Graph{f}}$ for the relational
$\Theta$-environment with $\myvec{\Graph{f}}_1 = \gamma$, and
$\myvec{\Graph{f}}_2 = \gamma'$ and $\myvec{\Graph{f}}_\R(\TVX) = \Graph{\myvec{f}_{\TVX}}$ 
and $\myvec{\Graph{f}}_\R(\algX) = \Graph{\myvec{f}_{\algX}}$.
Also, given a $\Theta$-environment, $\gamma$, we write $\myvec{x} \in \gamma$
for a family of elements indexed by type variables in $\Theta$ satisfying:
for every value-type variable $\TVX \in \Theta$, it holds that
$\myvec{x}_{\TVX} \in \gamma(\TVX)$ and,
for every computation-type variable $\algX \in \Theta$,
it holds that
$\myvec{x}_{\algX} \in U(\gamma(\algX))$.
Given a $\Theta$-environment morphism $\myvec{f} \colon \gamma \to \gamma'$ and
$\myvec{x} \in \gamma$, we write $\myvec{f(x)}$ for the evident 
pointwise function application, which is an element of $\gamma'$.
Moreover,
given a relational $\Theta$-environment $\rho$, and elements
$\myvec{x_1} \in \rho_1$ and $\myvec{x_2} \in \rho_2$,
we write $\rhoR(\myvec{x_1},\myvec{x_2})$ to mean that:
for every $\TVX \in \Theta$, it holds that
$\rhoR(\TVX)(\myvec{x_1}_{\TVX}, \myvec{x_2}_{\TVX})$; and,
for every $\algX \in \Theta$, it holds that
$\rhoR(\algX)(\myvec{x_1}_{\algX}, \myvec{x_2}_{\algX})$.

\begin{prop}[Groupoid action]
\label{prop:groupoidnew}
For any  type $\TU(\Theta)$,  any two 
$\Theta$-environments $\gamma$, $\gamma'$, 
and any $\Theta$-environment isomorphism $\myvec{i} \colon \gamma \to \gamma'$, 
there exists a unique isomorphism 
\[\Grpd{\TU}{\myvec{i}} \colon \SemC{\gamma}{\TU} \to \SemC{\gamma'}{\TU}\]
such that
\[
\SemCR{\myvec{\Graph{i}}}{\TU} \: = \: \Graph{\Grpd{\TU}{\myvec{i}}}\enspace .
\]
Moreover, if $\TU$ is a computation type then  $\Grpd{\TU}{\myvec{i}}$ is a homomorphism
from $\SemA{\gamma}{\TU}$ to $\SemA{\gamma'}{\TU}$.

Furthermore, given relational $\Theta$-environments $\rho$, $\rho'$, and 
given $\Theta$-environment isomorphisms $\myvec{i_1} \colon \rho_1 \to \rho'_1$ and
$\myvec{i_2} \colon \rho_2 \to \rho'_2$; if, for all $\myvec{x_1} \in \rho_1, \myvec{x_2} \in \rho_2$,
\[
\rho_R(\myvec{x_1}, \myvec{x_2}) \: \implies \: \rho'_R (\myvec{i_1(x_1)}, \myvec{i_2(x_2)}) \enspace ,
\]
then, for all ${x_1} \in \SemC{\rho_1}{\TU}, {x_2} \in \SemC{\rho_2}{\TU}$, we have:
\[
\SemCR{\rho}{\TU}(x_1,x_2) \: \implies \: 
\SemCR{\rho'}{\TU}(\Grpd{\TU}{\myvec{i_1}}(x_1), \Grpd{\TU}{\myvec{i_2}}(x_2)) \enspace .
\]
\end{prop}

\proof
By induction on the structure of the type $\TU$. We consider two cases.

If $\TU$ is $\TS \to \TT$ then the induction hypothesis gives
isomorphisms \mbox{$\Grpd{\TS}{\myvec{i}} \colon \SemC{\gamma}{\TS} \to \SemC{\gamma'}{\TS}$} and
$\Grpd{\TT}{\myvec{i}} \colon \SemC{\gamma}{\TT} \to \SemC{\gamma'}{\TT}$. Using that
$\SemCR{\myvec{\Graph{i}}}{\TS}  =  \Graph{\Grpd{\TS}{\myvec{i}}}$ and
$\SemCR{\myvec{\Graph{i}}}{\TT}  =  \Graph{\Grpd{\TT}{\myvec{i}}}$,
one calculates that 
\[\SemCR{\myvec{\Graph{i}}}{\TS \to \TT} \: = \: 
\Graph{f \mapsto \Grpd{\TT}{\myvec{i}} \circ f \circ (\Grpd{\TS}{\myvec{i}})^{-1}} \enspace,\]
so we have:
\[\Grpd{\TS \to \TT}{\myvec{i}} \: = \: f \mapsto \Grpd{\TT}{\myvec{i}} \circ f \circ (\Grpd{\TS}{\myvec{i}})^{-1} \enspace,\]
which obviously is an isomorphism.
Further, $\TU$ is a computation type just when $\TT$ is, in which case we must show that $\Grpd{\TS \to \TT}{\myvec{i}}$, as defined above, is a homomorphism. By definition $\SemA{\gamma'}{\TS \to \TT}$ is a $\SemC{\gamma'}{\TS}$-fold product of $\SemA{\gamma'}{\TT}$ by itself as taken in $\CatA$, and each evaluation map $\ev_x$, for $x\in \SemC{\gamma'}{\TS}$, is a projection. It suffices to show that for each $x\in \SemC{\gamma'}{\TS}$ the composite $\ev_x \circ \Grpd{\TS \to \TT}{\myvec{i}}$ is a homomorphism. But
\begin{align*}
\ev_x \circ \Grpd{\TS \to \TT}{\myvec{i}} (f) & = \Grpd{\TT}{\myvec{i}} \circ f \circ (\Grpd{\TS}{\myvec{i}})^{-1} (x) \\
& =  \Grpd{\TT}{\myvec{i}} \circ \ev_{(\Grpd{\TS}{\myvec{i}})^{-1} (x)} (f)
\end{align*}
and $\Grpd{\TT}{\myvec{i}}$ is a homomorphism by induction hypothesis, and evaluation maps are homomorphisms because they are projections out of a product taken in $\CatA$.

For the second half of the proposition, given isomorphisms $\myvec{i_1} \colon \rho_1 \to \rho'_1$ and
$\myvec{i_2} \colon \rho_2 \to \rho'_2$ as in the hypothesis, we must show that if $\SemCR{\rho}{\TS \to \TT}(f_1, f_2)$ and $\SemCR{\rho'}{\TS}(x_1, x_2)$ then 
\begin{equation} \label{eq:rel:preservation}
\SemCR{\rho'}{\TT}(\Grpd{\TT}{\myvec{i_1}} \circ f_1 \circ (\Grpd{\TS}{\myvec{i_1}})^{-1}(x_1), \Grpd{\TT}{\myvec{i_2}} \circ f_2 \circ (\Grpd{\TS}{\myvec{i_2}})^{-1}(x_2))
\end{equation}
Note first that $(\Grpd{\TS}{\myvec{i_1}})^{-1} = \Grpd{\TS}{\myvec{i_1}^{-1}}$ because
\begin{align*}
\Graph{(\Grpd{\TS}{\myvec{i_1}})^{-1}} & = \oprel{\Graph{\Grpd{\TS}{\myvec{i_1}}}} \\
& = \oprel{\SemCR{\myvec{\Graph{i_1}}}{\TS}} \\
& = \SemCR{\oprel{\myvec{\Graph{i_1}}}}{\TS} \\
& = \SemCR{\Graph{\myvec{i_1}^{-1}}}{\TS} \\
& = \Graph{\Grpd{\TS}{\myvec{i_1}^{-1}}}
\end{align*}
where we have used Lemma~\ref{lemma:opposite}. Similarly $(\Grpd{\TS}{\myvec{i_2}})^{-1} = \Grpd{\TS}{\myvec{i_2}^{-1}}$. So by the induction hypothesis, under the assumptions stated above 
\[
\SemCR{\rho}{\TS}((\Grpd{\TS}{\myvec{i_1}})^{-1}(x_1), (\Grpd{\TS}{\myvec{i_2}})^{-1}(x_2))
\]
and so also 
\[
\SemCR{\rho}{\TT}(f_1 \circ (\Grpd{\TS}{\myvec{i_1}})^{-1}(x_1), f_2 \circ (\Grpd{\TS}{\myvec{i_2}})^{-1}(x_2))
\]
from which we conclude (\ref{eq:rel:preservation}) by a second application of the induction hypothesis.

We define $\Grpd{\All{\algX}{\TT}}{\myvec{i}}$ by the formula
\[(\Grpd{\All{\algX}{\TT}}{\myvec{i}}(\kappa))_{\algA} = \Grpd{\TT}{\myvec{i}[\id_{\algA} / \algX]} (\kappa_{\algA})\enspace .
\]
to see that this is well defined we must show that if $\algA, \algC \in \SetA$ and $Q \in \RelA{\algA,\algC}$ then
\begin{equation} \label{eq:gpd:All:welldef}
\SemCR{\Delta_{\gamma'}[Q/\algX]}{\TT}(\Grpd{\TT}{\myvec{i}[\id_{\algA} / \algX]}(\kappa_{\algA}), \Grpd{\TT}{\myvec{i}[\id_{\algC} / \algX]}(\kappa_{\algC})) \enspace .
\end{equation}
But since $\SemCR{\Delta_{\gamma}[Q/\algX]}{\TT}(\kappa_{\algA}, \kappa_{\algC})$ and since the pair $(\myvec{i}[\id_{\algA}], \myvec{i}[\id_{\algC}])$ maps pairs related in $\Delta_{\gamma}[Q/\algX]$ to pairs related in $\Delta_{\gamma'}[Q/\algX]$ the induction hypothesis implies (\ref{eq:gpd:All:welldef}).

To show $\SemCR{\Graph{\myvec{i}}}{\All{\algX}{\TT}} = \Graph{\Grpd{\All{\algX}{\TT}}{\myvec{i}}}$, first suppose that  $\SemCR{\myvec{\Graph{i}}}{\All{\algX}{\TT}}(\kappa_1, \kappa_2)$. Then $\SemCR{\Graph{\myvec{i}[\id_{\algA} / \algX]}}{\TT}((\kappa_1)_{\algA}, (\kappa_2)_{\algA})$ for all $\algA$ and so by induction hypothesis $((\kappa_1)_{\algA}, (\kappa_2)_{\algA})$ is in $\Graph{\Grpd{\TT}{\myvec{i}[\id_{\algA} / \algX]}}$ which implies $\Grpd{\All{\algX}{\TT}}{\myvec{i}}(\kappa_1) = \kappa_2$. Suppose on the other hand that $\Grpd{\All{\algX}{\TT}}{\myvec{i}}(\kappa_1) = \kappa_2$. Then 
\[\SemCR{\Delta_{\gamma'}[Q/\algX]}{\TT}((\Grpd{\All{\algX}{\TT}}{\myvec{i}}(\kappa_1))_{\algA}, (\kappa_2)_{\algC})
\]
for all $\algA, \algC \in \SetA$ and $Q \in \RelA{\algA,\algC}$, i.e., 
\begin{equation} \label{eq:myeq}
\SemCR{\Delta_{\gamma'}[Q/\algX]}{\TT}(\Grpd{\TT}{\myvec{i}[\id_{\algA}/\algX]}((\kappa_1)_{\algA}), (\kappa_2)_{\algC}) \enspace .
\end{equation}
The pair $(\myvec{i}^{-1}[\id_{\algC}/\algX], \id_{\gamma'}[\id_{\algA}/\algX])$ maps pairs related in $\Delta_{\gamma'}[Q/\algX]$ to pairs related in $\Graph{\myvec{i}}[Q/\algX]$, and so by induction hypothesis,  the pair $(\Grpd{\TT}{\myvec{i}^{-1}[\id_{\algA}/\algX]}, \Grpd{\TT}{\id_{\gamma'}[\id_{\algA}/\algX]})$ maps pairs related in $\SemCR{\Delta_{\gamma'}[Q/\algX]}{\TT}$ to pairs related in $\SemCR{\Graph{\myvec{i}}[Q/\algX]}{\TT}$. As above, one can show that 
\[\Grpd{\TT}{\myvec{i}^{-1}[\id_{\algA}/\algX]} = (\Grpd{\TT}{\myvec{i}[\id_{\algA}/\algX]})^{-1}\]
and using Lemma~\ref{lemma:opposite} also $\Grpd{\TT}{\id_{\gamma'}[\id_{\algA}/\algX]} = \id_{\SemC{\gamma'[\algA/\algX]}{\TT}}$  and so by (\ref{eq:myeq}) we conclude 
\[\SemCR{\Graph{\myvec{i}}[Q/\algX]}{\TT}((\kappa_1)_{\algA}, (\kappa_2)_{\algC}) \enspace .
\]
Since this holds for all $\algA, \algC \in \SetA$ and $Q \in \RelA{\algA,\algC}$ this implies $\SemCR{\Graph{\myvec{i}}}{\All{\algX}{\TT}}(\kappa_1, \kappa_2)$. In conclusion we have shown $\SemCR{\Graph{\myvec{i}}}{\All{\algX}{\TT}} = \Graph{\Grpd{\All{\algX}{\TT}}{\myvec{i}}}$.

The type $\All{\algX}{\TT}$ is a computation type exactly when $\TT$ is, and in this case we must show that $\Grpd{\All{\algX}{\TT}}{\myvec{i}}$ is a homomorphism. Similarly to the case of function spaces, since $\SemA{\gamma'}{\All{\algX}{\TT}}$ is constructed as a limit in $\CatA$ it suffices to show that each composite $\proj{\algA} \circ \Grpd{\All{\algX}{\TT}}{\myvec{i}}$ is a homomorphism, where $\proj{\algA}$ is the projection defined as $\proj{\algA}(\kappa) = \kappa_{\algA}$. Since 
\begin{align*}
\proj{\algA} \circ \Grpd{\All{\algX}{\TT}}{\myvec{i}}(\kappa) & =  \Grpd{\TT}{\myvec{i}[\id_{\algA}/\algX]} (\kappa_{\algA}) \\
& =  \Grpd{\TT}{\myvec{i}[\id_{\algA}/\algX]} \circ \proj{\algA} (\kappa) 
\end{align*}
this follows by the induction hypothesis.

For the last part of the proposition, suppose the pair $(\myvec{i_1}, \myvec{i_2})$ maps pairs related in $\rho$ to pairs related in $\rho'$, and suppose $\SemCR{\rho}{\All{\algX}{\TT}}(\kappa_1, \kappa_2)$. We must show that 
\[\SemCR{\rho'}{\All{\algX}{\TT}}(\Grpd{\All{\algX}{\TT}}{\myvec{i_1}}(\kappa_1), \Grpd{\All{\algX}{\TT}}{\myvec{i_2}}(\kappa_2)) \enspace,\]
i.e., we must show that for any $\algA, \algC \in \SetA$, $Q\in \RelA{\algA, \algC}$
\[\SemCR{\rho'[Q/\algX]}{\TT}(\Grpd{\TT}{\myvec{i_1}[\id_{\algA}/\algX]}((\kappa_1)_{\algA}), \Grpd{\TT}{\myvec{i_2}[\id_{\algC}/\algX]}((\kappa_2)_{\algC})) \enspace .\]
Since the pair $(\myvec{i_1}[\id_{\algA}/\algX], \myvec{i_2}[\id_{\algC}/\algX])$ maps pairs related in $\rho[Q/\algX]$ to pairs related in $\rho'[Q/\algX]$ this follows from the induction hypothesis.
\qed

\begin{cor} 
The mapping of isomorphisms between $\Theta$-environments, $\myvec{i}$, to
$\Grpd{\TU}{\myvec{i}}$ is functorial.
\end{cor}
\proof
Preservation of identities is Lemma~\ref{lemma:identity}. For preservation
of composition, suppose $\myvec{i} \colon \rho \to \rho'$ and $\myvec{j} \colon \rho' \to \rho''$. 
If $\Graph{\myvec{i}}(\myvec{x}, \myvec{y})$ then 
$\Graph{\myvec{j} \circ \myvec{i}}(\myvec{x}, \myvec{j}(\myvec{y}))$ so by Proposition~\ref{prop:groupoidnew}, if
$\SemCR{\Graph{\myvec{i}}}{\TU}(x,y)$ then $\SemCR{\Graph{\myvec{j} \circ \myvec{i}}}{\TU}(x, \Grpd{\TU}{\myvec{j}}(y))$. Since $\SemCR{\Graph{\myvec{i}}}{\TU}(x, \Grpd{\TU}{\myvec{i}}(x))$, we conclude 
\[\SemCR{\Graph{\myvec{j} \circ \myvec{i}}}{\TU}(x, \Grpd{\TU}{\myvec{j}}\circ \Grpd{\TU}{\myvec{i}} (x))\]
for all $x$, i.e.,  $\Grpd{\TU}{\myvec{j}}\circ \Grpd{\TU}{\myvec{i}} = \Grpd{\TU}{\myvec{j} \circ \myvec{i}}$ as desired.
\qed

\begin{cor} \label{cor:reindexing}
For any type $\TT(\Theta, \TVX)$, relational $\Theta$-environment $\rho$, any relation $R$ in $\RelC{\SA,\SC}$, and any pair of isomorphisms $i\colon \SA' \to \SA$, $j\colon \SC' \to \SC$
\[\SemCR{\rho[\inv{(i,j)}R/\TVX]}{\TT} = \inv{(\Gpd{\rho_1}{\TT}{i}{\TVX}, \Gpd{\rho_2}{\TT}j{\TVX})}  \SemCR{\rho[R/\TVX]}{\TT} \enspace .
\]
Similarly for any type $\TT(\Theta, \algX)$, relational $\Theta$-environment $\rho$, any relation $R \in \RelA{\algA,\algC}$, and any pair of isomorphisms $i\colon \algA' \lfun \algA$, $j\colon \algC' \to \algC$
\[\SemCR{\rho[\inv{(i,j)}R/\algX]}{\TT} = \inv{(\Gpd{\rho_1}{\TT}{i}{\algX}, \Gpd{\rho_2}{\TT}j{\algX})}  \SemCR{\rho[R/\algX]}{\TT} \enspace .
\]
\end{cor}

\proof
We just prove the first part. Since the pair $(i,j)$ maps pairs related in $\inv{(i,j)}R$ to pairs related in $R$, by Proposition~\ref{prop:groupoidnew} the pair $(\Gpd{\rho_1}{\TT}{i}{\TVX}, \Gpd{\rho_2}{\TT}j{\TVX})$ maps pairs related in $\SemCR{\rho[\inv{(i,j)}R/\TVX]}{\TT}$  to pairs related in $\SemCR{\rho[R/\TVX]}{\TT}$. This means that 
\begin{equation} \label{eq:reindex:cor1}
\SemCR{\rho[\inv{(i,j)}R/\TVX]}{\TT} \subseteq \inv{(\Gpd{\rho_1}{\TT}{i}{\TVX}, \Gpd{\rho_2}{\TT}j{\TVX})}  \SemCR{\rho[R/\TVX]}{\TT} \enspace .
\end{equation}
Since $R = \inv{(\inv{i}, \inv{j})} \inv{(i, j)} R$ we can apply the above to the pair $(\inv{i}, \inv{j})$ and obtain 
\begin{align*}
\SemCR{\rho[R/\TVX]}{\TT}  \subseteq \inv{(\Gpd{\rho_1}{\TT}{\inv{i}}{\TVX}, \Gpd{\rho_2}{\TT}{\inv{j}}{\TVX})}  \SemCR{\rho[\inv{(i,j)}R/\TVX]}{\TT} 
\end{align*}
from which we conclude
\begin{align} \label{eq:reindex:cor2}
\inv{(\Gpd{\rho_1}{\TT}{i}{\TVX}, \Gpd{\rho_2}{\TT}j{\TVX})}  \SemCR{\rho[R/\TVX]}{\TT} \subseteq 
 \SemCR{\rho[\inv{(i,j)}R/\TVX]}{\TT} \enspace .
\end{align}
The corollary is now the collected statement of (\ref{eq:reindex:cor1}) and (\ref{eq:reindex:cor2}).
\qed

Now, for  any set $\SA$ in $\CatC$, let
$\SC \in \SetC$ be such that $\SC \cong \SA$ by way of the isomorphism
$i \colon \SC \to \SA$. Using the groupoid action defined above,
we have $\Gpd{\gamma}{\TT}{i}{\TVX}(\pi_{\SC}) \in \SemC{\gamma[\SA/\TVX]}{\TT}$.
Similarly, for any algebra $\algA$ in $\CatA$, let 
$\algC \in \SetA$ be such that
$\algC \algiso \algA\,$
by way of
$j \colon \algC \lfun \algA$. Then we have
$\Gpd{\gamma}{\TT}{j}{\algX}(\kappa_{\algC}) \in \SemC{\gamma[\algA/\algX]}{\TT}$.
\begin{lem} 
\label{lemma:invariant}
For $\pi \in \SemC{\gamma}{\All{\TVX}{\TT}}$ and $\SA$ in $\CatC$:
\begin{enumerate}[\em(1)]
\item The value $\Gpd{\gamma}{\TT}{i}{\TVX}(\pi_{\SC})$ is independent of the choice of $\SC$ and $i$.
\item If $\SA \in \SetC$ then $\Gpd{\gamma}{\TT}{i}{\TVX}(\pi_{\SC}) = \pi_{\SA}$.
\end{enumerate}
Similarly, for $\kappa \in \SemC{\gamma}{\All{\algX}{\TT}}$ and $\algA \in \CatA$:
\begin{enumerate}[\em(1)]
\setcounter{enumi}{2}
\item The value $\Gpd{\gamma}{\TT}{j}{\algX}(\kappa_{\algC})$ is independent of the choice of $\algC$ and $j$.
\item If $\algA \in \SetA$ then $\Gpd{\gamma}{\TT}{j}{\algX}(\kappa_{\algC}) = \kappa_{\algA}$.
\end{enumerate}
\end{lem}
\proof
We prove 1. Suppose $i\colon \SC \to \SA, i' \colon \SC'\to \SA$ are isomorphisms. We must show that $\Gpd{\gamma}{\TT}{i}{\TVX}(\pi_{\SC}) = \Gpd{\gamma}{\TT}{i'}{\TVX}(\pi_{\SC'})$. By the parametricity condition in the definition of $\SemC{\gamma}{\All{\TVX}{\TT}}$, $\SemC{\Delta_{\gamma}[\Graph{\inv{i'}\circ i} / \TVX]}{\TT}(\pi_{\SC}, \pi_{\SC'})$, which means that 
\[\Graph{\Gpd{\gamma}{\TT}{\inv {i'} \circ i}{\TVX}} (\pi_{\SC}, \pi_{\SC'}) \, .\]
Now by definition of graph relations and functoriality of the groupoid action this implies \[\Gpd{\gamma}{\TT}{i}{\TVX}(\pi_{\SC}) = \Gpd{\gamma}{\TT}{i'}{\TVX}(\pi_{\SC'})\] as desired.
Item 2 is an immediate consequence: use the identity on $\SA$ for $i$.
\qed

The above lemma justifies introducing the following very useful
notation.  Given $\pi$ in $\SemC{\gamma}{\All{\TVX}{\TT}}$, then, for
any $\SA$ in $\CatC$, we write $\pi(\SA)$ for
$\Gpd{\gamma}{\TT}{i}{\TVX}(\pi_{\SC})$, where $i \colon \SC \to \SA$
is an isomorphism and $\SC \in \SetC$. Similarly, given $\kappa \in
\SemC{\gamma}{\All{\algX}{\TT}}$, then, for any $\algA \in \CatA$, we
write $\kappa(\algA)$ for
$\Gpd{\gamma}{\TT}{j}{\algX}(\kappa_{\algC})$, where $j \colon \algC
\lfun \algA$ is an isomorphism and $\algC \in \SetA$.  The above
notation defines the required projections exhibiting $\pi$ and
$\kappa$ as elements of large products indexed by the objects of
$\CatC$ and $\CatA$ respectively. The lemma below shows that the
tuples $\pi$ and $\kappa$ remain parametric when considered as
elements of the large products, i.e., that the derived projections
preserve relations.\footnote{In the conference version of this
paper~\cite{mogelberg:simpson:lics07}, we saved space by using
fictitious large products in the definition of the interpretation of
polymorphic types. Here, by 
giving the honest definition, and
deriving the required consequences, we are providing the
missing technical justification for the use of large products 
in \emph{op.\ cit.}}    

\begin{lem} 
\label{lemma:useful}
\leavevmode
\begin{enumerate}[\em(1)]
\item If  $\SemCR{\rho}{\All{\TVX}{\TT}}\, (\pi,\pi')$ then, for all sets
$\SA,\SC$ in $\CatC$ and relations $R \in \RelC{\SA,\SC}$, it holds that 
$\SemCR{\rho[R/\TVX]}{\TT}(\pi(\SA),\pi'(\SC))$.
\item If  $\SemCR{\rho}{\All{\algX}{\TT}}\, (\kappa,\kappa')$ then, for all algebras
$\algA,\algC$ in $\CatA$ and relations $Q \in \RelA{\algA,\algC}$, it holds that 
$\SemCR{\rho[Q/\algX]}{\TT}(\kappa(\algA),\kappa'(\algC))$.
\end{enumerate}
\end{lem}

\proof
We just prove item 1 of the lemma, item 2 is proved similarly. Suppose we are given sets $\SA,\SC$ in $\CatC$ and a relation $R \in \RelC{\SA,\SC}$. Then we know that there exists sets $\SA', \SC' \in \SetC$ and isomorphisms $i\colon \SA' \to \SA$, $j\colon \SC' \to \SC$. By definition, if $\SemCR{\rho}{\All{\TVX}{\TT}}\, (\pi,\pi')$ then $\SemCR{\rho[\inv{(i,j)}R/\TVX]}{\TT}\,(\pi_{\SA'}, \pi'_{\SC'})$ and so by Corollary~\ref{cor:reindexing} 
\[\inv{(\Gpd{\rho_1}{\TT}i{\TVX},\Gpd{\rho_2}{\TT}j{\TVX})}\SemCR{\rho[R/\TVX]}{\TT}\,(\pi_{\SA'}, \pi'_{\SC'}) \,. \] 
So $(\pi(\SA), \pi'(\SC)) = (\Gpd{\rho_1}{\TT}{\inv{i}}{\TVX}(\pi_{\SA'}),\Gpd{\rho_2}{\TT}{\inv{j}}{\TVX}(\pi'_{\SC'}))$ are in $\SemCR{\rho[R/\TVX]}{\TT}$.
\qed

\begin{lem} \label{lemma:proj:hom}
For any computation type $\TB(\Theta, \TVX)$, any $\Theta$ environment $\gamma$ and any $\SA$ in $\CatC$ the projection $\proj{\SA} \colon \SemA{\gamma}{\All{\TVX}{\TB}} \to \SemA{\gamma[\SA/\TVX]}{\TB}$ mapping $\kappa$ to $\kappa(\SA)$ is a homomorphism. Similarly for any $\TB(\Theta, \algX)$ and any $\algA \in \CatA$ the projection $\proj{\algA} \colon \SemA{\gamma}{\All{\algX}{\TB}} \to \SemA{\gamma[\algA/\algX]}{\TB}$ is a homomorphism.
\end{lem}
\proof
Note first that for $\SA$ in $\SetC$, the projection $\proj{\SA}$ is a homomorphism since $\SemA{\gamma}{\All{\TVX}{\TB}}$ is defined as a representative of an $\CatA$-subobject of a $\SetC$ indexed $\CatA$-product and $\proj{\SA}$ is the inclusion of the subobject followed by the projection. In general,  $\proj{\SA}(\kappa)$ is defined to be
\[\Gpd{\gamma}{\TB}{i}{\TVX} (\proj{\SA'}(\kappa))
\]
for any $\SA'\in \SetC$, and isomorphism $i\colon \SA' \to \SA$. Since by Proposition~\ref{prop:groupoidnew} $\Gpd{\gamma}{\algB}{i}{\TVX}$ is a homomorphism, we see that $\proj{\SA}$ is a composition of homomorphisms and so itself a homomorphism. The second half of the lemma is proved similarly.
\qed

Next, we define the interpretation of terms.  Given a context $\Gamma$
with all free type variables in $\Theta$, a
$\Theta$-$\Gamma$-environment is a function defined on both the type
variables in $\Theta$ and the term variables in $\Gamma$, such that
the restriction of $\gamma$ to $\Theta$ is a $\Theta$-environment,
and, for every type assigment $\In{x}{\TT}$ in $\Gamma$, it holds that
$\gamma(x) \in \SemC{\gamma}{\TT}$.  A term
$\ajth{\Gamma}{\Delta}{t}{\TT}{\Theta}$ (i.e., such that
$\ftv{\Gamma,\Delta,t,\TT} \subseteq \Theta$) is interpreted as an
element $\Sems{\gamma}{t} \in \SemC{\gamma}{\TT}$, relative to any
$\Theta$-$(\Gamma,\Delta)$-environment $\gamma$.  The definition of
$\Sems{\gamma}{t}$ is given in Figure~\ref{figure:terms}.
In the two clauses that apply to  $t(\TA)$, we distinguish between
the cases for $t$ of type $\All{\TVX}{\TT}$ and $\All{\algX}{\TT}$.
Note that the definition of $\Sems{\gamma}{s(t)}$ applies uniformly, whether $s$ has type
$\TT \to \TU$ or $\TA \lfun \TB$.

\begin{figure}
\begin{align*}
\Sems{\gamma}{x} & = \gamma(x) \\
\Sems{\gamma}{\lam{x}{\TT}{t}}  =  \Sems{\gamma}{\llam{x}{\TA}{t}} & = (d \colon \SemC{\gamma}{\TT} \mapsto \Sems{\gamma[d/x]}{t}) \\
\Sems{\gamma}{s(t)} & = \Sems{\gamma}{s}(\Sems{\gamma}{t}) \\
\Sems{\gamma}{\Lam{\TVX}{t}} & =  \{ \Sems{\gamma[\SA/\TVX]}{t} \}_{\SA \in \SetC} \\
\Sems{\gamma}{{t[\colon \All{\TVX}{\TT}]}(\TS)} & =  (\Sems{\gamma}{t})({\SemC{\gamma}{\TS}}) \\
\Sems{\gamma}{\Lam{\algX}{t}} & =  \{ \Sems{\gamma[\algA/\algX]}{t} \}_{\algA \in \SetA} \\
\Sems{\gamma}{{t[\colon \All{\algX}{\TT}]}(\TA)} & =  (\Sems{\gamma}{t})({\SemA{\gamma}{\TA}})
\end{align*}
\vspace*{-.7cm}
\caption{Interpretation of Terms}
\label{figure:terms}
\end{figure}

\begin{prop} 
\label{propossition:interpret}
If $\ajth{\Gamma}{\Delta}{t}{\TT}{\Theta}$ then:
\begin{enumerate}[\em(1)]
\item \label{itA} (Well-definedness)
For any $\Theta$-$(\Gamma,\Delta)$-environment $\gamma$,
the value $\Sems{\gamma}{t} \in \SemC{\gamma}{\TT}$ is well defined.

\item \label{itC} (Relational invariance)
For any 
relational $\Theta$-environment $\rho$, and
$\Theta$-$(\Gamma,\Delta)$-environments $\gamma_1, \gamma_2$ extending $\rho_1, \rho_2$ respectively,
define
\begin{align*}
& \SemCR{\rho}{\Gamma}(\gamma_1, \gamma_2) \Iff 
\forall {\In{x}{\TS}  \in (\Gamma,\Delta)}. \, \SemCR{\rho}{\TS}(\gamma_1(x), \gamma_2(x)) .
\end{align*}
Then  $\SemCR{\rho}{\Gamma}(\gamma_1, \gamma_2)$ implies 
$\SemCR{\rho}{\TT}(\Sems{\gamma_1}{t},\Sems{\gamma_2}{t})$.
\end{enumerate}
If $\ajth{\Gamma}{\In{x}{\TA}}{t}{\TB}{\Theta}$ then:
\begin{enumerate}[\em(1)]
\setcounter{enumi}{2}
\item \label{itB} (Homomorphism property)
For any $\Theta$-$\Gamma$-environment $\gamma$,
the function
$d \in \SemC{\gamma}{\TA}  \mapsto \Sems{\gamma[d/x]}{t}$ 
is a homomorphism
from $\SemA{\gamma}{\TA}$ to $\SemA{\gamma}{\TB}$.
\end{enumerate}
\end{prop}

\proof[Proof (sketch).]
The three statements of the proposition are proved simultaneously by structural induction on $t$. Most of the cases are standard and we just show a few.

We prove the homomorphism property in the case of application of a polymorphic term $t\colon \All{\TVX}{\algB}$ to a value type $\TS$. By definition
\[\Sems{\gamma}{{t}(\TS)} = \proj{\SemC{\gamma}{\TS}}(\Sems{\gamma}{t})
\]
and so by the induction hypothesis and Lemma~\ref{lemma:proj:hom} $d \mapsto \Sems{\gamma[d/x]}{{t}(\TS)}$ is a composition of homomorphisms.

The homomorphism property in the case of function application  $t(s)$ for $t\colon \TB \lfun \TC$ follows from well definedness: by induction hypothesis $\Sems{\gamma}{t} \in \SemC{\gamma}{\TB \lfun \TC}$ and so is a homomorphism, so if $d \in \SemC{\gamma}{\TA}  \mapsto \Sems{\gamma[d/x]}{s}$  is a homomorphism so is $d \in \SemC{\gamma}{\TA}  \mapsto \Sems{\gamma}{t}(\Sems{\gamma[d/x]}{s})$. Likewise well definedness in the case of linear lambda abstraction: $\llam{x}{\TA}{t}$ follows from the homomorphism property for $t$. 

We show well definedness in one of the cases of polymorphic lambda abstraction: $\Lam{\TVX}{t}\colon \All{\TVX}{\TT}$. Here we must show that $\{ \Sems{\gamma[\SA/\TVX]}{t} \}_{\SA \in \SetC}$ satisfies the parametricity condition in the definition of $\SemC{\gamma}{\All{\TVX}{\TT}}$: for all $\SA, \SB\in \SetC$ and all relations $R\in \RelC{\SA, \SB}$, 
\[\SemCR{\Delta_{\gamma}[R/ \TVX]}{\TT}(\Sems{\gamma[\SA/\TVX]}{t}, \Sems{\gamma[\SB/\TVX]}{t})
\]
This follows from the relational invariance property for $t$, as assumed in the induction hypothesis, since $\SemCR{\Delta_{\gamma}}{\Gamma}(\gamma, \gamma)$ holds by the identity extension lemma. 
Likewise, the relational invariance property in the case of type application of polymorphic terms follows from well definedness using Lemma~\ref{lemma:useful}.

To show relational invariance in case of polymorphic application at computation types $t(\TA)$ we may use the induction hypothesis
\[\SemCR{\rho}{\All{\algX}{\TT}}(\Sems{\gamma_1}{t}, \Sems{\gamma_2}{t}).
\]
From Lemma~\ref{lemma:useful} it follows that 
\[\SemCR{\rho[\SemCR{\rho}{\TA}/\algX]}{\TT}(\Sems{\gamma_1}{t} (\SemA{\gamma_1}{\TA}), \Sems{\gamma_2}{t}(\SemA{\gamma_2}{\TA})).
\]
Finally, Lemma~\ref{lemma:substitution} implies 
\[\SemCR{\rho}{\TT[\TA/\algX]}(\Sems{\gamma_1}{t (\TA)}, \Sems{\gamma_2}{t(\TA)})
\]
as desired.
\qed

Our main application of  the model will be to establish semantic equalities between
terms. 
Henceforth, for $\aj{\Gamma}{\Delta}{s}{\TT}$ 
and $\aj{\Gamma}{\Delta}{t}{\TT}$, we write
$\aj{\Gamma}{\Delta}{s \; = \; t}{\TT}$ to mean
that $\Sems{\gamma}{s} = \Sems{\gamma}{t}$ for 
all appropriate $\gamma$. For a syntactic equality theory we refer to~\cite{mogelberg:simpson:logic}.

\section{Monadic types}
\label{section:monadic}

In this section, we study the encoding of monadic
types $\bang{\TT}$ in our calculus, as defined by
equation~(\ref{equation:central}) of
Section~\ref{section:introduction}. One sees immediately
that $\bang{\TT}$ is always a computation type. 
We show that it enjoys the following derived 
introduction and elimination rules.
\[
\prooftree
\tj{\Gamma}{t}{\TT}
\justifies 
\tj{\Gamma}{\bang{t}}{\bang{\TT}}
\endprooftree
\GAP
\prooftree
\aj{\Gamma}{\Delta}{t}{\bang{\TT}}
\GAP
\tj{\Gamma,\, \In{x}{\TT}}{u}{\TA}
\justifies
\aj{\Gamma}{\Delta}{\Let{x}{t}{u}}{\TA}
\endprooftree
\]
Indeed, for this simply define:
\begin{align*}
\bang{t} \, & \eqdef \, \Lam{\algX}{\lam{p}{\TT \to \algX}{p(t)}} \\
\Let{x}{t}{u} \, & \eqdef \, t(\TA)(\lam{x}{\TT}{u}) \enspace .
\end{align*}
It is the above rules that motivate our notation
for the $\bang{}$ type constructor, since these are 
simply restrictions of the usual rules for the exponential $\bang{}$
of intuitionistic linear logic; for example, 
as formulated in Plotkin and Barber's DILL~\cite{barber:phd}.

As a first application of relational parametricity for our system, 
we show that $\bang{\TT}$ has the correct universal property for 
Moggi's monadic type. 
To keep the semantic notation bearable, we frequently
omit semantic brackets, treating syntactic objects as the
semantic elements they define, and we freely mix
syntactic expressions with semantic values.
For example, given any set $\SA$ in $\CatC$, we simply write
$\bang{\!\SA}$ rather than $\SemC{[\SA/\TVX]}{\bang{\TVX}}$
or $\SemA{[\SA/\TVX]}{\bang{\TVX}}$, referring 
to $\bang{\!\SA}$ as a set or as an algebra respectively
when disambiguation 
is needed.

\begin{lem}
\label{lemma:bang}
\leavevmode
\begin{enumerate}[\em(1)]
\item \label{bang:beta}
If $\tj{\Gamma}{t}{\TT}$ and $\tj{\Gamma,\, \In{x}{\TT}}{u}{\TA}$
then $\tj{\Gamma}{\Let{x}{\bang{t}}{u} \; = \; u[t/x]}{\TA} \,$.

\item \label{bang:eta}
$\aj{\Gamma}{\In{y}{\bang{\TS}}}{y \; = \; \Let{\!x}{y}{\bang{\!x}}}{\bang{\TS}}\, $.

\item \label{bang:kappa}
Suppose that $\aj{\Gamma}{\Delta}{\!s}{\bang{\TS}}$,
$\;\;\tj{\Gamma  ,  \In{x}{{\TS}}}{\!t}{\TB}$ and
$\aj{\Gamma}{\In{y}{\TB}}{u}{\TC}$, then
$\aj{\Gamma}{\Delta}{\Let{\!x}{s}{u[t/y]} \; = \; u[\, \Let{\!x}{s}{t} \, / \, y]}{\TC}\,$.
\end{enumerate}
\end{lem}

\proof
Item~\ref{bang:beta} is a straightforward consequence of the semantic
validity of beta equality.

For~\ref{bang:eta},
we must show that $y = y(\bang{\TS})(\lam{x}{\TS}{\bang{x}})$ at type
$\All{\algX}{(\TS \to \algX) \to \algX}$. By evident extensionality properties
of the model, it suffices to show that, for any
algebra $\algB$ and $f \colon {\TS} \to U\algB$ in $\CatC$, we have
$y(\algB)(f) = y(\bang{\TS})(\lam{x}{\TS}{\bang{x}})(\algB)(f)$.

Consider the homomorphism $g \colon \bang{\TS} \lfun \algB$ defined by
$g(z) \! = \! z(\algB)(f)$. Then $\Graph{g}$ is in $\RelA{\bang{\TS},\algB}$.
So, by parametricity, 
\begin{equation}
\label{rel:a}
((\Delta_{\TS} \to \Graph{g}) \to \Graph{g}) \: (y(\bang{\TS}), \, y(\algB)) \enspace .
\end{equation}
For any $x \in \TS$, we have
$g(\bang{\!x}) = (\Lam{\algX}{\lamnt{p}{p(x)}})(\algB)(f) = f(x)$, i.e.,
\begin{equation}
\label{rel:b}
(\Delta_{\TS} \to \Graph{g})\:(\lam{x}{\TS}{\bang{\!x}},\, f) \enspace .
\end{equation}
Combining~(\ref{rel:a}) and~(\ref{rel:b}), we obtain that
\[
\Graph{g} \: (y(\bang{\TS})(\lam{x}{\TS}{\bang{\!x}}), \, y(\algB)(f)) \enspace ,
\]
i.e., $g(y(\bang{\TS})(\lam{x}{\TS}{\bang{\!x}})) = y(\algB)(f)$.
Thus it indeed holds that 
\[y(\bang{\TS})(\lam{x}{\TS}{\bang{x}})(\algB)(f) = y(\algB)(f)\,.\]

For~\ref{bang:kappa}, $h = \llam{y}{\TB}{u} \colon \TB \lfun \TC$ is
a homomorphism, so $\Graph{h} \in \RelA{\TB,\TC}$.
By parametricity, we have that
\begin{equation}
\label{rel:c}
((\Delta_{\TS} \to \Graph{h}) \to \Graph{h}) \: (s(\TB), \, s(\TC)) \enspace .
\end{equation}
Consider $\lam{x}{\TS}{t} \colon \TS \to \TB$ and
$\lam{x}{\TS}{u[t/y]} \colon \TS \to \TC$. Then,
for $x \in \TS$, it holds that
$h((\lam{x}{\TS}{t})(x)) = u[t/y] = (\lam{x}{\TS}{u[t/y]})(x)$,
i.e.,
\begin{equation}
\label{rel:d}
(\Delta_{\TS} \to \Graph{h}) \: (\lam{x}{\TS}{t}, \, \lam{x}{\TS}{u[t/y]}) \enspace .
\end{equation}
Combining~(\ref{rel:c}) and~(\ref{rel:d}), we obtain that
\[
\Graph{h} \: (s(\TB)(\lam{x}{\TS}{t}), \, s(\TC)(\lam{x}{\TS}{u[t/y]})) \enspace ,
\]
i.e., $h(s(\TB)(\lam{x}{\TS}{t})) = s(\TC)(\lam{x}{\TS}{u[t/y]})$.
So indeed we have
$u[\, \Let{\!x}{s}{t} \, / \, y] = h(s(\TB)(\lam{x}{\TS}{t})) = s(\TC)(\lam{x}{\TS}{u[t/y]}) =
\Let{\!x}{s}{u[t/y]}$.
\qed

Lemma~\ref{lemma:bang}  can be formulated as the two equality rules 
for the monadic type let constructor.
\[
\prooftree
\tj{\Gamma}{t}{\TT}
\GAP
\tj{\Gamma,\, \In{x}{\TT}}{u}{\TA}
\justifies
\aj{\Gamma}{-}{\Let{x}{\bang{t}}{u} \; = \; u[t/x]}{\TA}
\endprooftree
\GAP \GAP
\prooftree
\aj{\Gamma}{\Delta}{s}{\bang{\TS}}
\GAP
\aj{\Gamma}{\In{y}{{\bang\TS}}}{\!u}{\TC}
\justifies
\aj{\Gamma}{\Delta}{\Let{\!x}{s}{u[\bang x/y]} \; = \; u[s \, / \, y]}{\TC}
\endprooftree
\]
It is not hard to show that the two rules above are equivalent to
the three items of Lemma~\ref{lemma:bang}
and we leave this as a straightforward exercise.

For any set $\SA$ in $\CatC$ define $\eta_\SA \colon \SA \to \bang{\!\SA}$
by $\eta_\SA = \lamnt{x}{\bang{\!x}}$.

\begin{thm} \label{thm:free:alg}
The function $\eta_\SA \colon \SA \to \bang{\!\SA}$ presents $\bang{\!\SA}$ as the free algebra over $\SA$, i.e., for 
any algebra $\algB$ and function  $f \colon \SA \to U\algB$, there
exists a unique homomorphism $h \colon \bang{\!\SA} \lfun \algB$ such that
$h \circ \eta_\SA = f$. Indeed, $h$ is given by
$\llamnt{y}{\Let{\!x}{y}{f(x)}}$.
\end{thm}
\proof
Clearly $\llamnt{y}{\Let{\!x}{y}{f(x)}}$ is a homomorphism, and
$(\llamnt{y}{\Let{\!x}{y}{f(x)}}) \circ \eta_\SA = f$ because
$\Let{\!x}{\bang{\!x}}{f(x)} = f(x)$ by 
Lemma~\ref{lemma:bang}.\ref{bang:beta}.
For uniqueness, suppose $h$ is such that 
$h \circ \eta_\SA = f$. Then
\begin{align*}
h(y) & = h(\Let{x}{y}{\bang{\!x}}) & & \text{(Lemma~\ref{lemma:bang}.\ref{bang:eta})} \\
 & = \Let{x}{y}{h(\bang{\!x}}) & & \text{(Lemma~\ref{lemma:bang}.\ref{bang:kappa})} \\
 & = \Let{x}{y}{f(x)}  & & \text{($h \circ \eta_\SA = f$)} \enspace ,
\end{align*}
as required.
\qed
It follows from the above theorem that the operation mapping $\SA$ to
the algebra $\bang{\!\SA}$ is the object part of a functor $F\colon \CatC \to \CatA$
left adjoint to $U$. We write $T$ for the associated monad $UF$ on $\CatC$.

The bijective correspondence of Theorem~\ref{thm:free:alg} can be expressed in 
the type theory $\PE$ as an isomorphism of (value) types between
$\bang{\!\TS} \lfun \TB$ and $\TS \to \TB$ given by terms
\begin{align*}
\lam{f}{\TS \to \TB}{\llam{z}{\bang{\!\TS}}{\Let{\!x}{z}{f(x)}}} &\colon (\TS \to \TB) \to \bang{\!\TS} \lfun \TB \\
\lam{g}{\bang{\!\TS} \lfun \TB}{\lam{x}{\TS}{g(\eta_{\TS}(x))}} &\colon (\bang{\!\TS} \lfun \TB) \to  \TS \to \TB \; .
\end{align*}
Thus we have a Girard decomposition
of function spaces with computation type codomains, further motivating
the $\bang{}$ notation.


We end this section with three characterisations of the 
induced relational lifting of the $\bang{}$ type constructor.

\begin{prop}\label{prop:bang:rel}
Suppose $\SA,\SB$ are objects of $\CatC$ and $R\in\RelC{\SA,\SB}$ is a relation. 
\begin{enumerate}[\em(1)]
\item \label{item:bang:rel:1} $\bang R\in \RelA{\bang \SA,\bang \SB}$ is the smallest admissible $\CatA$-relation containing 
all pairs of the form $(\eta(x), \eta(y))$ for $(x,y)\in R$.
\item \label{item:bang:rel:2} $\bang R$ is the smallest admissible relation containing the image of the map \mbox{$T R \to T \SA \times T \SB$} obtained by applying the functor $T$ to the span corresponding to $R$.
\item \label{item:bang:rel:3} If $\algA, \algB \in \CatA$, $R\in \RelC{\SA, \SB}$, $Q\in \RelA{\algA, \algB}$ and $f\colon \bang\SA \lfun \algA, g\colon \bang \SB \lfun \algB$, then $(\bang R \lfun Q)(f,g)$  iff $(R \to Q)(f \circ \eta_{\SA},g \circ \eta_{\SB})$.
\end{enumerate}
\end{prop}

\proof
For item~\ref{item:bang:rel:1} we first show that if $(x,y)\in R$ then $(\eta_{\SA}(x), \eta_{\SB}(y)) \in \bang R$. So suppose we are given $\algA, \algB\in \CatA$ and $Q\in \RelA{\algA, \algB}$. We must show that if $f\colon \SA \to U \algA, g\colon \SB \to U\algB$  satisfy $(R \to Q)(f,g)$ then $Q(\eta_{\SA}(x)(\algA)(f), \eta_{\SB}(y)(\algB)(g))$. But this follows from definition of $(R \to Q)$ since $(\eta_{\SA}(x)(\algA)(f), \eta_{\SB}(y)(\algB)(g) = (f(x), g(y))$. 

Now, suppose $Q\in \RelA{\bang \SA, \bang \SB}$  and for all $(x,y) \in R$ we have $Q(\eta_{\SA} (x), \eta_{\SB} (y))$, or in other words $(R \to Q)(\eta_{\SA},\eta_{\SB})$. We must show that $\bang R \subseteq Q$. So suppose $\bang R(z,z')$.  By definition of $\bang R$ using $(R \to Q)(\eta_{\SA},\eta_{\SB})$ we have
\[Q(z (\bang \SA)(\eta_{\SA}), z' (\bang \SB)(\eta_{\SB})).
\]
But by definition $z (\bang \SA)(\eta_{\SA}) = \Let{x}{z}{\bang x}$ which by Lemma~\ref{lemma:bang} is equal to $z$. Likewise $z' (\bang \SB)(\eta_{\SB}) = z'$ proving $Q(z,z')$.

For the proof of item~\ref{item:bang:rel:2} we use the notation $\admcl{\im(T R)}$ for the smallest admissible relation containing the image of the map obtained by applying $T$ to the span corresponding to $R$. Since $(R \to \admcl{\im(T R)})(\eta_{\SA}, \eta_{\SB})$, by item~\ref{item:bang:rel:1}  the relation $\bang R$ is contained in $\admcl{\im(T R)}$. For the other inclusion notice that since $(R \to \bang R)(\eta_{\SA}, \eta_{\SB})$, 
naturality of the correspondence given by Theorem~\ref{thm:free:alg} implies the existence of a map $h$ making the diagram
\[
\begin{diagram}
T \SA \!\! & \lMulti^{T{\pi_1}} & T R & \rMulti^{T{\pi_2}} & T{\SB} \\
\dEq & & \dMulti_{h} & & \dEq \\
T{\SA} & \lMulti & \bang R & \rMulti & T{\SB}
\end{diagram}
\]
commute. This proves $\im(T R) \subseteq \bang R$. Since $\bang R$ is admissible $\admcl{\im(T R)}$ must be contained in $\bang R$. 

For item~\ref{item:bang:rel:3} the "only if" direction is simply because $(R\to \bang R)(\eta_{\SA}, \eta_{\SB})$. On the other hand, if $(R \to Q)(f \circ \eta_{\SA},g \circ \eta_{\SB})$ then $\inv{(f,g)}Q$ is an admissible relation containing all elements of the form $(\eta_{\SA}(x), \eta_{\SB}(y))$ for which $R(x,y)$ hold, and so by item~\ref{item:bang:rel:1} must contain $\bang R$ proving $(\bang R \lfun Q)(f,g)$.
\qed

\section{Definable computation types}
\label{section:def:computation:types}

\begin{figure}
\begin{align*}
\algone  & \eqdef  \All{\algX}{0 \to \algX} \\
\TA \algtimes \TB  & \eqdef  \All{\algX}{((\TA \! \lfun \! \algX) + (\TB \! \lfun \! \algX)) \to \algX} \!\!
 & & \!\!\text{($\algX \! \not\in \! \ftv{\TA,\!\TB}$)} \\
\algzero  & \eqdef  \All{\algX}{\algX} \\
\TA \algplus \TB  & \eqdef  \All{\algX}{(\TA \! \lfun \! \algX) \to (\TB \! \lfun \! \algX) \to \algX} \!\!
 & & \!\!\text{($\algX \!\not\in\! \ftv{\TA,\!\TB}$)} \\
\TT \! \cdot \TA  & \eqdef  \All{\algX}{(\TT \to \TA \lfun \algX) \to \algX} 
 & & \!\!\text{($\algX \!\not\in\! \ftv{\TT,\!\TA}$)} \\
\algexists{\TVX}{\TA}  & \eqdef  \All{\algY}{(\All{\TVX}{(\TA \lfun \algY})) \to \algY}  
 & & \!\!\text{($\algY \!\not\in \!\ftv{\TA}$)} \\
\algexists{\algX}{\TA} & \eqdef  \All{\algY}{(\All{\algX}{(\TA \lfun \algY})) \to \algY}  
 & & \!\!\text{($\algY \!\not\in\! \ftv{\TA}$)} \\
\algMu{\algX}{\TA}  & \eqdef  \All{\algX}{(\TA \lfun \algX) \to \algX} & & \text{($\algX$ +ve in $\TA$)} \\
\algNu{\algX}{\TA}  & \eqdef  \algexists{\algX}{(\algX \lfun \TA) \! \cdot \algX} & & \text{($\algX$ +ve in $\TA$)} 
\end{align*}
\vspace*{-.7cm}
\caption{Definable computation types}
\label{figure:definable-algebra}
\end{figure}

The monadic type constructor $\bang{}$ is just one example of a type constructor definable using parametric polymorphism. In Figure~\ref{figure:definable} we have seen a collection of type constructors on value types and Figure~\ref{figure:definable-algebra} presents a collection of type constructors on computation types. The latter should be viewed as well chosen variants of Plotkin's polymorphic type encodings in second-order intuitionistic linear type theory, cf.~\cite{plotkin:type-theory-recursion,bpr:lily,Mogelberg:lapl:journal}. (For relations between this calculus and $\PE$ see Section~\ref{section:derived}). 
We briefly discuss the computation type encodings.

Semantically, because $U \colon \CatA \to \CatC$ weakly creates limits,
algebras are closed under products in $\CatC$. Syntactically, however,
the types $\one$ and  $\TA \times \TB$ from Figure~\ref{figure:definable} are \emph{not}
computation types. Thus the alternative encodings $\algone$ and $\TA \algtimes \TB$ are needed 
to obtain products of computation types as  computation types. 
The types $\algzero$ and $\TA \algplus \TB$
from Figure~\ref{figure:definable-algebra} define respectively
an initial object and binary coproduct in the category $\CatA$.
This structure in \emph{not} preserved by $U$,
and coproducts of algebras behave very differently
from coproducts of sets in $\CatC$. (The latter are 
implemented by the sum types in Figure~\ref{figure:definable}.)
The type $\TT \! \cdot \TA$ defines
a ${\SemC{}{\TT}}$-fold copower of $\SemA{}{\TA}$ in
$\CatA$. Figure~\ref{figure:definable-algebra} also contains:
existential types, $\algexists{\TVX}{\TA}$ and $\algexists{\algX}{\TA}$, packaged up as computation types;
inductive computation types, $\algMu{\algX}{\TA}$; and
coinductive computation types, $\algNu{\algX}{\TA}$. 
As is standard, the (co)inductive types rely on the functoriality of type expressions
in their positive arguments.
A special case of the inductive types is the isomorphism
\begin{align*}
\TA  & \algiso  \All{\algX}{(\TA \lfun \algX) \to \algX}  
\end{align*}
valid for all computation types $\TA$ in which  $\algX$ does not occur free.
It is a consequence of relational parametricity
that the above types all enjoy the correct universal properties. 
The arguments are carried out most naturally using a 
suitable logic for relational parametricity in $\PE$, 
see~\cite{mogelberg:simpson:logic}.

\section{Specialising the calculus to specific effects}
\label{section:special}

The type theory $\PE$ is a generic calculus for effects since
the type $\bang{\TT}$ can be interpreted as an arbitrary monad, and
no further effect-specific features are included. In this regard,
$\PE$ is analogous to Moggi's computational 
$\lambda$-calculus~\cite{Moggi:89}, 
computational metalanguage~\cite{moggi:monads-journal} and
Levy's call-by-push-value~\cite{levy:cbpv}. As with those
calculi, specific effects can be incorporated by
specialising the calculus appropriately. Typically, such
specialisation takes place by extending the basic calculus with
appropriately typed constants for any desired operations 
on effects. The addition of such constants takes place within
the semantic theory described thus far, and so does not
affect the validity of the results we have presented. For example,
the universal properties of the defined types, discussed
in Sections~\ref{section:monadic} and~\ref{section:def:computation:types}
(and treated in more detail in~\cite{mogelberg:simpson:logic}), are
unaltered.

In this 
section we consider various specialisations of the basic calculus,
emphasising, in particular, the interaction
with parametricity.

In a recent programme of research~\cite{plotkin-power:overview},
Plotkin and Power have shown that many
monads of computational interest can
be profitably viewed as free algebra constructions
for equational theories. This approach 
arises naturally from a computational viewpoint: the ``algebraic operations''
used to specify the theory correspond to programming
primitives that cause effects, and the equational theory
simply expresses natural 
behavioural equivalences between such primitives.
We begin this section with an analysis of
how to specialise $\PE$ to the case of such 
``algebraic effects''. 

Our approach is justified by a general 
theorem, which we now present. As one of their 
central results about algebraic effects, 
Plotkin and Power
establish a one-to-one correspondence between 
``algebraic operations'' and ``generic effects''~\cite{plotkin-power:generic}. 
The theorem below reformulates this correspondence in our setting,
and adds a third equivalent induced by our polymorphic
description of monadic types. We shall apply this third equivalent
to obtain the correct polymorphic typing
for algebraic operations in 
effect-specific specialisations of $\PE$.

\begin{thm}
\label{theorem:generic}
For any set $\SA$ in $\CatC$, there are one-to-one correspondences between:
\begin{enumerate}[\em(1)]
\item \label{generic:b} 
``algebraic operations of arity $\SA$'', i.e., natural transformations from the functor \linebreak[4] $(U(-))^\SA \colon \CatA \to \CatC$ to $U$,

\item \label{generic:a} 
``generic effects over $\SA$'', i.e., elements of $T\SA$, and

\item  \label{generic:c} ``polymorphic computation type operations of
arity $\SA$'', that is,
       elements of the type \mbox{$\All{\algX}{(\SA \to \algX) \to \algX}$}.
\end{enumerate}
\end{thm}
\noindent
The simplifications in the formulation
of statement~\ref{generic:b} above,
compared with~\cite{plotkin-power:generic}, are due to our set-theoretic setting,
which renders it unnecessary to consider issues relating to 
enrichment or tensorial strength.
Also note that, by statement~\ref{generic:a}, the other two statements,
in spite of appearances, depend only on the monad $T$ on $\CatC$, not on 
how it is resolved into an adjunction $F \dashv U \colon \CatA \to \CatC$.

\proof
The equivalence of statements~\ref{generic:a} and~\ref{generic:c} is immediate
from~(\ref{equation:central}), because $T\SA = \bang{\!\SA}$. So we establish
the equivalence of~\ref{generic:b} and~\ref{generic:c}. Suppose that
$\theta$ is a natural transformation from $(U(-))^\SA$ to $U$. We
show that the mapping $\algA \in \CatA \mapsto \lam{f}{\SA \to U\algA}{\,\theta_{\algA}(f)}$ is an element
of \mbox{$\All{\algX}{(\SA \to \algX) \to \algX}$}. Suppose $\algA,\algB \in \CatA$ and
$Q \in \RelA{\algA,\algB}$. We must show that if
$(\Delta_{\SA} \to Q)\,(f,g)$ then also $Q(\theta_{\algA}(f), \theta_{\algB}(g))$.
Since $Q$ is an $\CatA$ relation there exists a span $\algA \gets \algC \to \algB$ in $\CatA$ projected by $U$ to $U\algA \gets Q \to U \algB$, and so by naturality
the two squares below commute.
\[
\begin{diagram}
(U\algA)^\SA \!\! & \lTo^{\;\;(\pi_1)^{\SA}} & Q^{\SA} & \rTo^{(\pi_2)^{\SA}} & U\algB^{\SA} \\
\dTo^{\theta_{\algA}} & & \dTo_{\theta_{\algC}} & & \dTo_{\theta_{\algB}} \\
U\algA & \lTo_{\pi_1} & Q & \rTo_{\pi_2} & U\algB
\end{diagram}
\]
But this says that, for any $f,g$ with $Q(f(x),g(x))$ for all $x \in \SA$, it holds
that $Q(\theta_{\algA}(f), \theta_{\algB}(g))$, which is what we needed to show.
For the converse direction, suppose $\kappa$ is an element of 
\mbox{$\All{\algX}{(\SA \to \algX) \to \algX}$}.
Then $\theta_{\algA}(f)  = \kappa(\algA)(f)$ is the corresponding
algebraic operation. Verifying naturality is a routine use of graphs of homorphisms:
if $g \colon \algB \lfun \algC$ and $f \colon \SA \to \algB$ then by parametricity 
\[((\Delta_{\SA} \to \Graph g) \to \Graph g)(\kappa(\algB), \kappa(\algC))\; ,\] 
so since $(\Delta_{\SA} \to \Graph g)(f, g\circ f)$, also $\Graph g( \kappa(\algB)(f),  \kappa(\algC)(g\circ f))$, i.e., $g(\theta_{\algB}(f)) = \theta_{\algC}(g\circ f)$ proving naturality.
It is obvious that the two constructions are mutually inverse.
\qed

To illustrate how
Theorem~\ref{theorem:generic} informs the specialisation of $\PE$
to algebraic effects, we consider nondeterminism
as a typical example.
As in~\cite{plotkin-power:overview},
nondeterministic choice is naturally formulated using
a binary operation ``$\bor$'' satisfying the semilattice equations:
\[
x \,\bor\, x = x, \;\; x \,\bor\, y = y \,\bor\, x, \;\; x \,\bor\, (y \,\bor\, z) = (x\, \bor \,y) \,\bor \,z \enspace .
\]
Define the category $\CatAnd$ of ``nondeterministic algebras'' to have, 
as objects, structures $(\SA, \bor_{\SA})$ where
$\SA$ is a set in $\CatC$ and $\bor_{\SA} \colon \SA \times \SA \to \SA$
satisfies the semilattice equations, and, as morphisms
from $(\SA, \bor_{\SA})$ to $(\SB, \bor_{\SB})$, functions
from $\SA$ to $\SB$ that are homomorphisms with respect to
the ``$\bor$'' operations. It is easily verified that
the obvious forgetful functor $U \colon \CatAnd \to \CatC$ satisfies
conditions (A1)--(A4). 

Since the morphisms in $\CatAnd$ are homomorphisms, 
the operation mapping any nondeterministic algebra $(\SA, \bor_{\SA})$
to the function $\bor_{\SA} \colon \SA^2 \to \SA$ is an algebraic
operation of arity $2$ in the sense of 
statement~\ref{generic:b} of Theorem~\ref{theorem:generic}.
Thus, applying Theorem~\ref{theorem:generic} and currying, one obtains a 
corresponding polymorphic operation:
\[
\bor \, \colon \; \All{\algX}{\,\algX \to \algX \to \algX} \enspace .
\]
Accordingly, nondeterministic choice can be incorporated in $\PE$
by adding a constant "$\bor$", typed as above,  to the type theory.
This example illustrates  the general pattern for adding algebraic operations as 
polymorphic constants to our type theory, and readily adapts to
the algebraic operations associated with other algebraic effects.

A limitation of the notion of algebraic operation is that
there exist effect-specific programming primitives that  are not
algebraic operations. One well-known example of such a primitive is
exception  handling. Below, we show how exception handling
may also be incorporated within our approach as
a suitably typed polymorphic constant. The approach
is justified by a general theorem, giving another
instance of a coincidence between natural transformations and 
elements of polymorphic type. 

\begin{thm}
\label{theorem:lin-gen}
For any  $n  \in \mathbb{N}$, there are one-to-one correspondences between:
\begin{enumerate}[\em(1)]
\item \label{lin-gen:b} 
Natural transformations from $(F(-))^n \colon \CatC \to \CatA$ to $F \colon \CatC \to \CatA$, and

\item  \label{lin-gen:c} 
       elements of $\All{\TVX}{(n \to \bang{\!\TVX}) \lfun \bang{\!\TVX}}$,
\end{enumerate}
where, in statement \ref{lin-gen:c}, we write $n$ for the $n$-fold coproduct  type
$1 + \dots + 1$, as defined in Figure~\ref{figure:definable}.
\end{thm}

\proof 
  An element of $\All{\TVX}{(n\to \bang{X}) \lfun \bang X}$ gives for
  each $\SA\in \CatC$ a map \mbox{$(F{\SA})^n \lfun F{\SA}$}, and the
  naturality square for this family follows from the parametricity
  condition satisfied by elements of polymorphic type, applied to the
  graph of a function. The interesting part of this proof is to show
  that natural transformations satisfy the parametricity condition and
  thus define elements of \mbox{$\All{\TVX}{(n\to \bang{X}) \lfun \bang X}$}.

So suppose $(f_{\SA} \colon (F{\SA})^n \lfun F{\SA})_{\SA \in \CatC}$ is a natural transformation, and $\SA, \SB \in \CatC$ and \mbox{$R\in \RelC{\SA,\SB}$}. We must show that $((\bang{R})^n \lfun \bang{R})(f_{\SA}, f_{\SB})$. Naturality applied to the span $\SA \leftarrow R \to \SB$ gives us commutativity of 
\[
\begin{diagram}
(F \SA)^n \!\! & \lMulti^{(F{\pi_1})^n} & (F R)^n & \rMulti^{(F{\pi_2})^n} & (F{\SB})^n \\
\dMulti^{f_{\SB}} & & \dMulti_{f_{R}} & & \dMulti_{f_{\SB}} \\
F{\SA} & \lMulti^{F{\pi_1}} & F R & \rMulti^{F{\pi_2}} & F{\SB}
\end{diagram}
\]
Since $f_{\SA}$ and $f_{\SB}$  are homomorphisms, this implies 
\[(\admcl{\im{((T R)^n)}} \lfun \admcl{\im(T R)}) (Uf_{\SA}, Uf_{\SB})
\]
Now, one can easily check that $\admcl{\im{((T R)^n)}} = (\admcl{\im{(T R)}})^n$ and so 
\mbox{$((\bang{R})^n \lfun \bang{R})(Uf_{\SA}, Uf_{\SB})$} by Proposition~\ref{prop:bang:rel},
as desired.
\qed

We now consider exception handling in detail. We assume we have a set
$E$ of exceptions with decidable equality (i.e., for all $e,e' \in E$
either $e = e'$ or $e \neq e'$). We also assume (for simplicity) that $\CatC$ 
is closed under binary coproduct in $\Sets$ (this is consistent with the axioms for $\CatC$). We define the category $\CatAexc$
of ``exception algebras'' to have, as objects, structures $(\SA, \{\braise^e_{\SA}\}_{e \in E})$
where $\braise^e_{\SA} \in \SA$, and, as morphisms from 
$(\SA, \{\braise^e_{\SA}\}_{e \in E})$ to $(\SB, \{\braise^e_{\SB}\}_{e \in E})$,
functions
from $\SA$ to $\SB$ that map each $\braise^e_{\SA}$ to $\braise^e_{\SB}$.
Since the $\braise^e$ elements are algebraic constants (operations
of arity $0$), they can be added to $\PE$ as constants:
\[\braise^e \colon \All{\algX}{\algX} \enspace .\]
As is standard, the 
forgetful  functor from 
$\CatAexc$ to $\CatC$, has as its left adjoint
the functor $F$ mapping $\SA$ to 
the exception algebra $(\SA + E, \{\mathrm{inr}(e)\}_{e \in E})$. 
For an exception $e \in E$, the handling operation 
over $\SA$ is the function 
$\bhandle^e_{\SA} \colon (F(\SA))^2 \to F(\SA)$ defined by
\[
\bhandle^e_{\SA} (p,q)\;   = \; \left \{
\begin{array}{ll} p & \text{if $p \neq \mathrm{inr}(e)$} \\
                 q & \text{if $p = \mathrm{inr}(e)$} \enspace .
\end{array} \right .
\]
It is easily shown that this specifies a natural transformation from 
the functor $(F(-))^2 \colon \CatC \to \CatAexc$ to \mbox{$F \colon \CatC \to \CatAexc$}.
In particular, the component
$\bhandle^e_{\SA}$ of the natural transformation 
does lie in $\CatAexc$ because the interpretation of $\braise^e$ in
the exception algebra $F(\SA)^2$ is the pair $(\mathrm{inr}(e), \mathrm{inr}(e))$.
Thus, by Theorem~\ref{theorem:lin-gen}, exception handling can be incorporated in $\PE$ by adding
typed constants:
\[
\bhandle^e \colon \All{\TVX}{(2 \to \bang{\!\TVX}) \lfun \bang{\!\TVX}} \enspace .
\]
The main surprise with this typing is that exception handling is 
given a ``linear'' type. 
From this typing, one of course obtains an associated
term  of the less informative type
\mbox{$\All{\TVX}{(2 \to \bang{\!\TVX}) \to \bang{\!\TVX}}$},
which is isomorphic to the expected type
\mbox{$\All{\TVX}{\bang{\!\TVX} \to \bang{\!\TVX} \to \bang{\!\TVX}}$}.

Paul Levy (personal communication) has pointed out that the above
account of exception handling is not robust, in the sense that,
in the presence of effects other than exceptions, the linear
typing of $\bhandle^e$ above is not always correct.
In situations in which handling is non-linear, one 
would expect the non-linear typing 
\mbox{$\All{\TVX}{\bang{\!\TVX} \to \bang{\!\TVX} \to \bang{\!\TVX}}$}
to still be correct.
However, 
Theorem~\ref{theorem:lin-gen} is no longer applicable to establish
parametricity. 
It would thus be interesting to find a general argument, valid in the presence
of other effects,  for the parametricity of handling.

Both Theorems~\ref{theorem:generic} and~\ref{theorem:lin-gen}
relate elements of certain polymorphic types with
natural transformations between associated functors.
In fact, more generally, 
for types that determine functors, parametricity implies
naturality (cf.~\cite{plotkin-abadi:logic}). However, the exact correspondences
between natural transformations and parametric elements 
established above depend crucially on the precise
forms of types considered there.

The forms of $n$-ary operation considered in this section by no means
exhaust the collection of operations of interest from an effects perspective.
Control operators provide a particularly interesting class of examples
that do not fit into this format. 
We briefly discuss how $\PE$ can be specialised to control
at the end of Section~\ref{section:derived}.

\section{Relation to other systems}
\label{section:derived}

Several computational
effects of interest, including nontermination, nondeterminism, and probabilistic
choice, give rise to monads on $\CatC$ that are \emph{commutative}, 
cf.~\cite{moggi:monads-journal}.
The collection of models of $\PE$ in which $\CatA$ is the
category of algebras for a commutative monad $T$ is of special
interest since, for such monads, 
the set of homomorphisms $\algA \lfun \algB$ between algebras $\algA,\algB$ 
carries a canonical algebra structure which provides a
closed structure on the category $\CatA$.
For such models, it is thus natural to modify our type system
by including  \mbox{$\TA \lfun \TB$} as a computation type. Making
this adjustment, one obtains second-order 
intuitionistic linear type theory as the fragment of computation types:
\begin{equation}
\label{equation:sill}
\algX \mid \TA \lfun \TB \mid \TA \to \TB \mid \All{\algX}{\TA} \enspace .
\end{equation}
Thus we obtain a rich collection of models for the type theory proposed by
Plotkin as a foundation for combining polymorphism and 
recursion~\cite{plotkin:type-theory-recursion}.

A simple application of 
the polymorphic encodings in 
Figures~\ref{figure:definable} 
and~\ref{figure:definable-algebra} 
is to translate
Levy's CBPV calculus~\cite{levy:cbpv} into $\PE$.
For this, coproducts and products of value types
are translated using $+$ and $\times$ from Figure~\ref{figure:definable},
products of computation types are translated using $\algtimes$
from Figure~\ref{figure:definable-algebra}, Levy's
$F$ constructor is translated using $\bang{}$, and
$U$ is simply ignored.

One of the properties of Levy's CBPV calculus
is that its adjunction 
models~\cite{levy:stacks}
are not required to satisfy any properties analogous
to our  conditions
(A1) and (A2). In 
Sections~\ref{section:semantic-setting} and~\ref{section:type-interpretation},
we exploited (A1) to satisfy the requirement that
$U(\SemA{}{\TA}) = \SemC{}{\TA}$, and 
(A2) to obtain that relations in $\CatA$ can be viewed as
special relations in $\CatC$ (cf.\ Lemma~\ref{lemma:pr}), which
is crucial in interpreting 
$\SemCR{}{\TA}$ as an admissible $\CatA$-relation.
We comment, however, that it is possible to generalise
our account of relational parametricity to
models in which (A1) is weakened to the requirement that 
$\CatA$ be small-complete and $U$
preserve limits (which always holds in Levy's models
since $U$ is a right adjoint), and in which condition (A2) is dropped
altogether. For such models, condition (A1) can then be engineered
by changing $\CatA$ to an equivalent category, and adjusting
$U$ accordingly, as in~\cite{mogelberg:simpson:mfps07};
or, more naturally, the semantics can be adjusted, rather than
the category, so as
to obtain a specified isomorphism $U(\SemA{}{\TA}) \cong \SemC{}{\TA}$,
instead of an equality. Dropping condition (A2) causes a more significant
complication. In its absence, it seems necessary to define a special
relational semantics for computation types, rather than 
inheriting the relational semantics for computation types from that
for value types (as done in Section~\ref{section:type-interpretation}).
Moreover, while such an approach is natural, it does make the 
semantic definitions significantly more complicated. In this
paper, we have chosen to assume properties (A1) and (A2),
since we value the
convenience of  simplified semantic definitions (which are anyway
complicated enough as they are!) over the added generality of having
a wider class of models. 

Finally, we mention how the interesting case of control operators can
be accommodated within $\PE$.
This cannot be achieved
by following the general methods of Section~\ref{section:special}, 
since the 
continuations monad $R^{R^{(-)}}$ does not
arise naturally as the free algebra for an algebraic theory, 
and the control primitives associated with continuations
are not algebraic operations. 
Nevertheless, it turns out that  $\PE$ can be
usefully specialised to the case of control by adding 
a polymorphic constant of type (using the defined type $\algzero$ from
Figure~\ref{figure:definable-algebra}):
\[
\All{\algX}{((\algX \lfun \algzero) \to \algzero) \; \lfun \; \algX} \enspace ,
\]
acting as a pointwise inverse to the canonical element of 
type 
$\All{\algX}{\algX \lfun ((\algX \lfun \algzero) \to \algzero)}$.
The resulting theory is studied in detail in a 
companion article~\cite{mogelberg:simpson:mfps07}, 
where it is shown that 
Hasegawa's results on polymorphic definability
in the second-order $\lambda\mu$-calculus~\cite{Hasegawa:LMCS06}
fall out  as special cases of constructions from Figure~\ref{figure:definable-algebra}.

\section{Applicability of results}
\label{section:applicability}

We have given a semantic account of relational
parametricity in the presence of computational
effects. From our working perspective within
IZF, this is parametrized on being given categories
$\CatC$ and $\CatA$ and families of relations
$\RC$ and $\RA$, satisfying axioms (C1)--(C4),
(A1)--(A4) and (R1)--(R4). Moreover, 
Proposition~\ref{prop:alg-model}, shows that such
data can be obtained whenever one has a 
monad $T$ on a category $\CatC$ satisfying
(C1)--(C4).

To conclude the paper, we outline how this
theory might actually be applied to prove properties
of polymorphic programs with effects. 
Suppose we have some given polymorphic $\lambda$-calculus 
$\mathbf{L}$ with
a choice of effect-primitives as the programmming
language of interest. The basic idea is to formulate
both the operational and denotational semantics
of $\mathbf{L}$ within IZF. The operational
semantics is treated in the standard way, for
which the use of classical logic is inessential. The
denotational semantics is developed using
the assumption of a category $\CatC$ satisfying 
(C1)--(C4). The construction of $\CatA$ and 
$\RC$ and $\RA$ will depend upon the effects present in the language.
For (a simple) example, if the only effect is nondeterministic choice
then $T$ can be defined to be the free-semilattice functor
over $\CatC$, and the entire model is then obtained via
Proposition~\ref{prop:alg-model}.
For general effects, the construction of the model
will be more complex than this, especially in the presence of
recursion, cf.~\cite{rosolini-simpson:strictness}.
Indeed, there is need for a uniform theory of how to
build such models; some hints in this direction
appear in~\cite{simpson:MFPS07}.

Once one has both operational semantics and  model,
the next step is to prove, within IZF, a 
\emph{computational adequacy} result for the model,
implying that the model is sound with respect to
operational equivalence. In examples considered
hitherto, such proofs have been obtained by 
standard logical-relations-based 
methods~\cite{simpson:csl-adequacy,simpson:apal04,rosolini-simpson:strictness}.
They rely only on having some appropriate non-triviality property of $\CatC$
(for example, that the natural numbers is an object of 
$\CatC$~\cite{simpson:csl-adequacy}).

Computational adequacy allows one to transfer equational
properties  of the denotational semantics to the operational
semantics. However, the above development has taken place in
IZF, together with the assumption of a category $\CatC$ satisfying
(C1)--(C4). We can therefore infer operational
properties within this metatheory; but, of course, we want to
be sure that such properties are actually true in the real world.
The remaining step is to use a transfer property which allows
us to conclude exactly this. 

The transfer property is based on the existence of realizability
models of IZF which possess within them categories
$\CatC$ satisfying
(C1)--(C4) and containing the natural numbers as an object.
As already discussed in Section~\ref{section:semantic-setting},
such models derive from 
the work of Hyland \emph{et.\ al.}~on small-complete small categories~\cite{hyland:small-complete,hrr:discrete}.
Now, the relevant realizability models all enjoy the property
of being \emph{$\Pi^0_2$-absolute}, meaning that a 
$\Pi^0_2$-sentence holds in the model if and only if it is true
externally.
This implies that properties of operational equivalence
that are true in the model are indeed true in reality,
see~\cite{simpson:csl-adequacy,simpson:apal04,rosolini-simpson:strictness}
for related arguments. 

We have outlined a programme of how one can 
potentially use the theory of 
parametricity developed in this paper to derive
operational properties of programs.
It would be good to have examples of such applications 
worked out in computationally interesting cases. 

There is, of course, a significant drawback with the
intuitionistic-set-theory-based approach we have been
following. The mathematical overheads are considerable.
It seems likely that a more practical theory of
parametricity for effects should be achievable
using direct operational methods. 
We leave this as an
interesting direction for future research.
It is plausible that 
the denotational approach we have
been following in this paper might be useful in
informing the development
of such an operational theory.

\section*{Acknowledgements}

We are indebted to Masahito Hasegawa for 
first suggesting that the polymorphic definition of $\bang{\TT}$
given by 
(\ref{equation:central}) should be a general phenomenon
within a monad-based framework incorporating both
linear and continuation-passing settings as special cases.
We thank both  him and Paul Levy for very helpful discussions, and the
anonymous referees for useful suggestions.

\end{document}